\documentclass[twocolumn,twocolappendix,preprint]{aastex63}
\usepackage{amsmath}
\usepackage{amsfonts}
\usepackage{amssymb}
\usepackage{xcolor}
\usepackage{longtable}
\usepackage{makecell}
\usepackage{graphicx}
\graphicspath{{./Images/}}
\usepackage{hyperref}
\usepackage[T1]{fontenc}

\newcommand{\vb}{\vec{B}}
\newcommand{\vnet}{v_\mathrm{net}}
\newcommand{\vavg}{\langle | v | \rangle}
\newcommand{\rhigh}{R_{\mathrm{high}}}
\newcommand{\ipole}{{\tt ipole }}
\newcommand{\kharma}{{KHARMA }}

\defcitealias{EHTC_2019_1}{EHTC M87~I}
\defcitealias{EHTC_2019_2}{EHTC M87~II}
\defcitealias{EHTC_2019_3}{EHTC M87~III}
\defcitealias{EHTC_2019_4}{EHTC M87~IV}
\defcitealias{EHTC_2019_5}{EHTC M87~V}
\defcitealias{EHTC_2019_6}{EHTC M87~VI}
\defcitealias{EHTC_2021_7}{EHTC M87~VII}
\defcitealias{EHTC_2021_8}{EHTC M87~VIII}
\defcitealias{EHTC_2023_9}{EHTC M87~IX}

\defcitealias{EHTC_2022_1}{EHTC SgrA~I}
\defcitealias{EHTC_2022_2}{EHTC SgrA~II}
\defcitealias{EHTC_2022_3}{EHTC SgrA~III}
\defcitealias{EHTC_2022_4}{EHTC SgrA~IV}
\defcitealias{EHTC_2022_5}{EHTC SgrA~V}
\defcitealias{EHTC_2022_6}{EHTC SgrA~VI}

\begin{document}

\title{Circular Polarization of Simulated Images of Black Holes}

\author[0000-0002-2514-5965]{Abhishek~V.~Joshi}
\affiliation{Department of Physics, University of Illinois, 1110 West Green Street, Urbana, IL 61801, USA}
\affiliation{Illinois Center for Advanced Study of the Universe, 1110 West Green Street, Urbana, IL 61801, USA}

\author[0000-0002-0393-7734]{Ben~S.~Prather}
\affiliation{Department of Physics, University of Illinois, 1110 West Green Street, Urbana, IL 61801, USA}
\affiliation{Los Alamos National Lab, Los Alamos, NM, 87545, USA}

\author[0000-0001-6337-6126]{Chi-kwan Chan}
\affiliation{Steward Observatory and Department of Astronomy, University of Arizona, 933 N. Cherry Ave., Tucson, AZ 85721, USA}
\affiliation{Data Science Institute, University of Arizona, 1230 N. Cherry Ave., Tucson, AZ 85721, USA}
\affiliation{Program in Applied Mathematics, University of Arizona, 617 N. Santa Rita, Tucson, AZ 85721, USA}

\author[0000-0002-8635-4242]{Maciek Wielgus}
\affiliation{Max-Planck-Institut für Radioastronomie, Auf dem Hügel 69, D-53121 Bonn, Germany}

\author[0000-0002-0393-7734]{Charles~F.~Gammie}
\affiliation{Department of Physics, University of Illinois, 1110 West Green Street, Urbana, IL 61801, USA}
\affiliation{Illinois Center for Advanced Study of the Universe, 1110 West Green Street, Urbana, IL 61801, USA}
\affiliation{Department of Astronomy, University of Illinois, 1002 West Green Street, Urbana, IL 61801, USA}
\affiliation{National Center for Supercomputing Applications, 605 East Springfield Avenue, Champaign, IL 61820, USA}

\begin{abstract}

Models of the resolved Event Horizon Telescope (EHT) sources Sgr A* and M87* are constrained by observations at multiple wavelengths, resolutions, polarizations, and time cadences.  In this paper we compare unresolved circular polarization (CP) measurements to a library of models, where each model is characterized by a distribution of CP over time.  In the library we vary the spin of the black hole, the magnetic field strength at the horizon (i.e. both SANE and MAD models), the observer inclination, a parameter for the maximum ion-electron temperature ratio assuming a thermal plasma, and the direction of the magnetic field dipole moment.  We find that ALMA observations of Sgr A* are inconsistent with all edge-on ($i = 90^\circ$) models.  Restricting attention to the magnetically arrested disk (MAD) models favored by earlier EHT studies of Sgr A*, we find that only models with magnetic dipole moment pointing away from the observer are consistent with ALMA data.  We also note that in 26 of the 27 passing MAD models the accretion flow rotates clockwise on the sky.  We provide a table of the mean and standard deviation of the CP distributions for all model parameters along with their trends. 

\end{abstract}


\section{Introduction}

We investigate the origin of circular polarization using first-principles models of synchrotron emitting systems, and study the distribution of expected circular polarization across a set of models at varying spin, magnetization, and electron distribution functions. 

The 2017 Event Horizon Telescope (EHT) campaign produced total intensity images of the supermassive black hole at the center of M87 (hereafter M87*) and the Milky Way (hereafter Sgr~A*) at a resolution of $\sim 25\mu$as \citep[][hereafter EHTC M87~I--VI]{EHTC_2019_1,EHTC_2019_2,EHTC_2019_3,EHTC_2019_4,EHTC_2019_5,EHTC_2019_6} and \citep[][hereafter EHTC SgrA~I--VI]{EHTC_2022_1,EHTC_2022_2,EHTC_2022_3,EHTC_2022_4,EHTC_2022_5,EHTC_2022_6}. Both reconstructed images show a ring surrounding a central flux depression. The ring is produced by synchrotron emission from hot gas surrounding the black hole and the central depression corresponds to lines of sight that cross the event horizon (the black hole ``shadow'') 

EHT images have been interpreted by comparison to a library of numerical models \citepalias{EHTC_2019_5, EHTC_2022_5} in which spin, flow magnetization, source inclination, and electron distribution function are varied but the time-averaged 1.3\,mm flux density is held fixed and consistent with the April 2017 observations \citepalias[][\citealt{Wielgus2022lc}]{EHTC_2019_4}. In particular, plasma flow models were generated using general relativistic magnetohydrodynamics (GRMHD) simulations, and then ray-traced using a general relativistic radiative transfer scheme;  the modeling procedure is described in detail in \cite{wong_patoka_2022}.  The models predict time-dependent image structure in all four Stokes parameters at frequencies where scattering is unimportant, time-dependent unpolarized flux density across the electromagnetic spectrum, and jet power.  

For M87*, the model comparison exercise found a subset of library models that satisfied all available observational constraints.  The most discriminating observational constraint was a lower limit on jet power of $10^{42} \mathrm{erg} \sec^{-1}$.  The favored models were highly magnetized - so-called magnetically arrested disk (MAD) models \citep{igumenshchev03,narayan03} in which the magnetic flux through the horizon is large enough to episodically push aside the accreting plasma - and contained a population of relatively cool electrons \citepalias{EHTC_2019_5}.

For Sgr A*, 11 observational constraints were used in the model comparison exercise.  No models satisfied all constraints \citepalias{EHTC_2022_5}.  The most discriminating observational constraint was a measure of fractional variability in the 1.3mm Atacama Large Millimeter Array (ALMA) light curve; almost all models that failed this test were found to be too variable\textemdash most models failed with $p<0.01$ for two-sample K-S tests comparing distributions of the ratio of standard deviation to mean flux averaged over 3 hour timescales. An incomplete list of possible explanations for this variability ``crisis'' is provided in \citetalias{EHTC_2022_5}, including the possible presence of slowly varying, resolved-out features that would make the fractional variability in the ALMA light curve a lower limit on the variability of the compact source.  Setting aside the variability constraint the Sgr A* model comparison identified a set of models that passed all remaining constraints.  The favored models were MAD models that contain a population of relatively cool electrons.  

EHT also recorded linear and circular polarization (LP and CP) data in the 2017 campaign.  LP and CP model comparison studies of M87* \citep[][hereafter EHTC M87~VII--VIII]{EHTC_2021_7,EHTC_2021_8} included constraints applied to the models based on LP maps and limits to the total CP fraction.  Model LP maps are sensitive to the magnetic field configuration of M87* and are highly constraining.  Unresolved CP measurements, by contrast, exclude only a few models.

\cite{EHTC_2023_9} (hereafter EHTC M87~IX) analyzed CP in the 2017 EHT campaign data and found evidence for nonzero CP. The Stokes V resolved structure could not be constrained, however, in contrast to image reconstructions in Stokes I, Q, and U.   EHTC M87~IX placed an upper limit on the magnitude of resolved CP fraction of 3.7\%.  Consistent with results from LP model comparisons  \citepalias{EHTC_2021_8}, the CP constraints favor highly magnetized simulations with relatively cooler electrons in the disk and jet. 

ALMA has recorded polarization data in the 1mm and 3mm bands for Sgr~A*, M87* and other AGN (low-luminosity AGN, radio-loud AGN and blazars for which horizon scale images are not possible with the current EHT resolution). \citet{bower_alma_2018} (2016 observations), \citet{Goddi2021} and \citet{wielgus_orbital_2022,Wielgus2023} (2017 observations) present this ALMA data with detections or limits on the unresolved LP and CP of Sgr~A*. Other unresolved CP observations of Sgr~A* from the Submillimeter Array (SMA) at 3mm and other wavelengths are given in \citet{munoz_circular_2012} and references therein.

It is expected that at the observing wavelength of 1.3 mm, emission is produced by the synchrotron process, which is expected to be strongly linearly polarized. The polarization state is modified, however, by propagation through a warm, magnetized plasma, through Faraday rotation and conversion. The combination of emission and propagation effects are complicated, so numerical radiative transfer methods are essential in understanding the polarization of M87* and Sgr~A*.

CP structure can be used to understand the magnetic field structure of the accretion disk and field geometry of GRMHD models under certain conditions of observing angles, and optical and Faraday thicknesses \citep[][]{tsunetoe_polarization_2021, moscibrodzka_2021, ricarte_cp_2021}. CP may also be a useful probe of plasma composition, since CP emission and Faraday rotation is sensitive to the electron-positron pair content \citep{wilson_positrons_1997,wardle_electron-positron_1998,homan_polarization_3c279_2009,anantua_positron_2020,emami_positron_2021}.

Unresolved CP measurements, particularly the handedness (or sign) of CP have been hypothesized to indicate the sense of rotation of the disk \citep{enslin_2003} or the magnetic field configuration: the structure and dipole moment of the magnetic field \citep{beckert_falcke_2002}. With constant handedness across long timescales indicating a constancy in either of these two properties. This is especially interesting in the case of Sgr~A* where the sign of CP has been observed to be constantly negative across decades for 1.3mm (and larger wavelength) observations. In this paper we investigate these hypotheses by comparing observational data at 1.3mm to a library of simulated images. As we still do not have robust horizon scale CP images of M87* and Sgr~A*, it is useful to compare integrated fractional CP values obtained from General Relativistic Radiative Transfer (GRRT) images to the ALMA data, to see which models are consistent. 

The net CP is
\begin{equation}
    \vnet \equiv \frac{\int d^2x\, V(x,y)}{\int d^2x\, I(x,y)}
\end{equation}
where $x,y$ are coordinates on the sky.  Here $I, V$ are the Stokes I and V images convolved with the beam. 
The net CP fraction is accessible from ALMA observations, with an effective beam size of $1'' \sim 10^5 GM/c^2$. The GRMHD model library reliably reproduces emission out to $200 \mu$as $\sim 40 G M/c^2$ diameter of the source and to compare our simulations to observations, we assume that there is no significant emission between the two scales. In \citetalias{EHTC_2022_2}, comparisons of horizon scale baselines and short baselines such as ALMA-APEX ($100$\,mas $\sim 2\times 10^4 G M/c^2$) suggest that at least $\sim 90\%$ of the flux density (up to $100$mas) arises from the horizon scale emission.

We will also occasionally refer to the average absolute CP fraction
\begin{equation}
    \vavg \equiv \frac{\int d^2x\, |V|}{\int d^2x\, I}.
\end{equation}
Evidently measurement of $\vavg$ requires a resolved image of the source and depends on the beam size.   

The paper is organized as follows. In section \ref{sec:cp_physical_origin} we review the origin of CP in first-principles models in which the dominant emission mechanism is synchrotron emission from a relativistic thermal distribution of electrons. In section \ref{sec:num_model} we describe the numerical models used in the analysis along with the parameters that characterize the resulting model library. In section \ref{sec:cp_dist_one} we investigate the CP properties of a single GRMHD model highlighting properties that are generalizable across most of the library. In section \ref{subsec:cp_distributions} we present the full library of $\vnet$ distributions along with fits for their dependence on model parameters. Using unresolved VLBI data of Sgr~A* \citep[][]{bower_alma_2018,wielgus_orbital_2022}, we attempt to constrain our models of Sgr~A* in section \ref{sec:constraints}. In section \ref{sec:discussion} we present a discussion on the models including caveats, followed by a conclusion.

\section{Physical Origins of Circular Polarization}\label{sec:cp_physical_origin}

\subsection{Polarized Radiative Transfer Equation}

The time-independent radiative transfer equation, in flat space, along a ray labeled by $ s $, in the Stokes basis  $ (I,Q,U,V) $, is:
\begin{equation}\label{eq:GenRadTrans}
\frac { d } { d s } \left( \begin{array} { l } { I } \\ { Q } \\ { U } \\ { V } \end{array} \right) = \left( \begin{array} { c } { j _ { I } } \\ { j _ { Q } } \\ { j _ { U } } \\ { j _ { V } } \end{array} \right) - \left( \begin{array} { c c c c } { \alpha _ { I } } & { \alpha _ { Q } } & { \alpha _ { U } } & { \alpha _ { V } } \\ { \alpha _ { Q } } & { \alpha _ { I } } & { \rho _ { V } } & { - \rho _ { U } } \\ { \alpha _ { U } } & { - \rho _ { V } } & { \alpha _ { I } } & { \rho _ { Q } } \\ { \alpha _ { V } } & { \rho _ { U } } & { - \rho _ { Q } } & { \alpha _ { I } } \end{array} \right) \left( \begin{array} { l } { I } \\ { Q } \\ { U } \\ { V } \end{array} \right).
\end{equation}
Here $ j_S$, $\alpha_S$ and $\rho_S$ are the emission, absorption and Faraday rotation and conversion coefficients (``rotativities'') of component S of the Stokes vector. The coefficients depend on the field strength and direction, the energy distribution function of particles (electrons for synchrotron radiation from an electron-ion plasma), and the frame in which they are measured.  Scattering  processes such as Compton scattering are negligible at mm wavelength for M87* and Sgr A* and have been neglected.  Circular polarization is described by the V component of the Stokes vector.  We follow the IEEE convention for definitions of the sign of CP where $V>0$ is right handed circular polarization such that the electric field vector rotates in a right hand direction, at a fixed point in space, with the thumb pointing along the direction of propagation  \citep{hamaker_polconventions_1996}.

Separating out the Stokes V equation, 
\begin{equation}\label{eq:StokesV}
    \frac{d}{d s}V = j_V - \alpha_V I - \rho_U Q + \rho_Q U - \alpha_I V.
\end{equation}
Evidently Stokes V can be altered by direct, circularly polarized emission; polarization-specific  absorption (i.e. the plasma acts as a circular polarizing filter); Faraday conversion from linear polarization to circular; and polarization-nonspecific absorption.  Polarization-nonspecific absorption $\alpha_I$ does not change the fractional circular polarization $V/I$.   

\subsection{Origin of Circular Polarization}

In a spatially uniform magnetic field production of circular polarization can occur in three ways, best seen in the radiative transfer equation (Eq. \ref{eq:GenRadTrans}). Intrinsic emission ($j_V$), selective absorption of CP ($\alpha_V I$) or Faraday conversion of linearly polarized into circular polarized light ($\rho_{ Q }$ and $\rho_{ U }$ components that interconvert $U$ and $Q$ respectively, with $V$).

Using the radiative transfer equation in a homogeneous source (a ``one-zone'' model), one can estimate the linear and circular polarization fractions of the emergent radiation \citep[][]{jones_transfer_1977,pandya16}. For parameters appropriate to EHT sources a one-zone model produces a linear polarization fraction that is large compared to the circular polarization fraction, and the dominant production mechanism is Faraday conversion \citepalias[see Fig. 8][]{EHTC_2023_9}{}{}.  The one-zone model overproduces both linear and circular polarization, however - spatial inhomogeneities are important - and thus the case for Faraday conversion as the dominant source of CP in both simulations and observations cannot be made with one-zone models.  We show using an example model in section \ref{subsec:coef_contribution} that although Faraday conversion is usually the dominant mechanism for production of circular polarization, intrinsic circularly polarized emission makes a non-negligible, and sometimes dominant, contribution.

    

\subsection{Transfer Coefficients for Thermal Distribution}

We adopt a thermal (Maxwell\textendash J{\"u}ttner) electron energy distribution function.  This is motivated by the notion that the electron distribution function is likely to have an approximately thermal core extending up to Lorentz factor $\sim 30$ that produces mm emission (a hollow distribution, one in which $f(\vec{p})$ has a minimum at $\vec{p}=0$, would be kinetically unstable, see \citealt{penrose_instabilities_1960}), and the idea that any superthermal tail on the distribution must not overproduce near infrared emission \citepalias[section 4.2.3][]{EHTC_2022_5} i.e the tail is constrained by the IR-to-mm color.  

The emission coefficients are summarized in \citet{dexter16, pandya16, marszewski_coefficients_2021}, the absorption coefficients follow from Kirchoff's law, and the Faraday coefficients $\rho_S$ are given in \cite{pandya18}.

It is helpful in understanding the symmetries of the transfer equations to write out the transfer coefficients explicitly in the frame of the plasma.  As is conventional in this field, we work in a Stokes basis where Stokes U corresponds to linear polarization at $\pm 45^\circ$ to the projection of the magnetic field on the plane perpendicular to the wavevector.  Then the emissivity fits are (following \citealt{dexter16}):
\begin{equation}\label{eq:jI}
    j_{I}\left(\nu, \theta \right)=\frac{n e^{2} \nu}{2 \sqrt{3} c \Theta_{e}^{2}} I_{I}(x),
\end{equation}
\begin{equation}\label{eq:jQ}
    j_{Q}\left(\nu, \theta \right)=\frac{n e^{2} \nu}{2 \sqrt{3} c \Theta_{e}^{2}} I_{Q}(x),
\end{equation}
\begin{equation}
    j_{U} = 0,
\end{equation}
and
\begin{equation}\label{eq:jV}
    j_{V}\left(\nu, \theta \right)=\frac{2 n e^{2} \nu \cot(\theta)}{3 \sqrt{3} c \Theta_{e}^{2}} I_{V}(x).
\end{equation}
Here $n$ is the electron number density, $-e$ is the electron charge, $\nu$ is the photon frequency, $\theta$ is the angle between the wave-vector and magnetic field (sometimes called the observer angle), $c$ is the speed of light, $\Theta_e \equiv k_B T_e / (m c^2)$ is the dimensionless electron temperature and $x=\nu/\nu_c$ with $\nu_c = (3/2) \nu_B \sin(\theta) \Theta_e^2$, $\nu_B = e |B|/(2 \pi m c)$ (the cyclotron frequency) and $m$ is the electron mass.  The $I_S$ do not change sign under field reversal $\mathbf{B} \rightarrow - \mathbf{B}$. 

The absorptivities can be found from Kirchoff's law ($B_\nu$ is the black-body function):
\begin{equation}
    j_{S} = \alpha_{S} B_\nu
\end{equation}
The Faraday coefficients (fits) are:
\begin{equation}\label{eq:rhoQ}
    \rho_{Q}=\frac{n e^{2} \nu_{B}^{2} \sin^{2}\theta}{m c \nu^{3}} f(X)\left[\frac{K_{1}\left(\Theta_{e}^{-1}\right)}{K_{2}\left(\Theta_{e}^{-1}\right)}+6 \Theta_{e}\right]
\end{equation}
\begin{equation}
    \rho_{U} = 0
\end{equation}
\begin{equation}\label{eq:rhoV}
    \rho_{V}=\frac{2 n e^{2} \nu_{B}}{m c \nu^{2}} \frac{K_{0}\left(\Theta_{e}^{-1}\right)}{K_{2}\left(\Theta_{e}^{-1}\right)} \cos \theta g(X)
\end{equation}
where 
\begin{equation}\label{eq:X}
    X=\left(\frac{3}{2 \sqrt{2}} 10^{-3} \frac{\nu}{\nu_{c}}\right)^{-1/2}
\end{equation}
\begin{equation}
    g(X) = 1 - 0.11 \ln{(1 + 0.035 X)}
\end{equation}
and $K_0$, $K_2$ are modified Bessel functions of the second kind at orders 0 and 2 respectively. A field reversal transforms $\theta \rightarrow \pi - \theta$, which thus reverses the sign of $j_V$, $\alpha_V$ and $\rho_V$.

In a field-aligned Stokes basis $j_U = \alpha_U = \rho_U = 0$, and thus the Faraday conversion term in the transfer equation reduces to $+\rho_Q U$. Stokes U in the field-aligned Stokes basis is therefore required to produce Stokes V by Faraday conversion.  The Stokes U transfer equation reduces in the field-aligned basis, for a uniform plasma, to $dU/ds = \rho_V Q - \alpha_I U -\rho_Q V$, so Stokes V can be produced by Faraday rotation of Stokes Q followed by Faraday conversion to Stokes V.  We must add a term to the transfer equation, however, if we force the Stokes basis to be field aligned at each point on the ray.  This term captures the effect of rotation of the field through an angle $\psi$ in the plane perpendicular to the line of sight, which interconverts Stokes Q and U, with  $dU/ds = \ldots + 2 d\psi/ds Q$, $dQ/ds = \ldots - 2 d\psi/ds U$.  Restated, emission of linearly polarized light elsewhere along the line of sight produces Stokes U locally that can be Faraday converted to Stokes V.  In the end Faraday conversion acts on linearly polarized light produced by some combination of Faraday rotation and field line rotation (or ``twist''). \citet{ricarte_cp_2021} explore this effect in detail along with the resulting properties of the magnetic field that are apparent in Faraday conversion produced CP maps of GRMHD models. 

\subsection{Symmetries of the Coefficients and RTE}\label{subsec:symmetries_RTE}

Below we investigate the net circular polarization associated with  general relativistic magnetohydrodynamic (GRMHD) models. The models are turbulent and the circular polarization fluctuates in time.  We are interested in the distribution of net circular polarization $f(\vnet)$ for a GRMHD model with fixed time-averaged millimeter wavelength flux density, black hole spin, magnetization, inclination, and electron distribution function parameters.  

The GRMHD equations are invariant under magnetic field inversion $\vb \rightarrow -\vb$, but the radiative transfer equation is not because some of the transfer coefficients depend on the sign of $\vb$.  To fully sample $f$ then, we ought to  include field-inverted models.  Here we describe the symmetry of the transfer coefficients under field inversion, and its effect on the solution to the transfer equation.

Under field inversion, the handedness of electron orbits around the magnetic field lines change sign: an electron that orbits clockwise on the sky moves counterclockwise after field inversion. This change in handedness flips the sign of circular polarization for the emitted radiation.  This implies that $j_V$ and $\alpha_V$ change sign.  In addition $\rho_V$, the coefficient governing Faraday rotation also reversus sign under field inversion (see Eq. \ref{eq:jI}\textendash \ref{eq:rhoV}).  None of the other coefficients change sign.  

For a single layer of plasma with a uniform magnetic field and no background radiation (Stokes vector vanishes where the line of sight enters the plasma) we find, using the analytic solution of the radiative transfer equation \cite{deglinnocenti85}, that Stokes U changes sign when the field is inverted because the only source is Faraday rotation ($\rho_V$), which also changes sign.  Similarly, Stokes V changes sign since $j_V$ and Faraday-rotation-generated Stokes U change sign but $\rho_Q$ does not (see Eq. \ref{eq:StokesV}).  In sum, in the single layer model Stokes I and Q are invariant under field inversion and Stokes U and V change sign.

Any deviation from the single layer model destroys the symmetry of the Stokes vector under field inversion.  For example, polarized background radiation provides initial Stokes U that is symmetrically converted to Stokes V, but this is added to directly emitted circularly polarized radiation ($j_V$) that is antisymmetric.  Multiple layer models are not symmetric, since the Stokes Q generated in one layer (symmetric ) gets rotated into Stokes U in the next layer, where it can be Faraday converted (symmetric) to Stokes V, and this is added to the antisymmetric direct emission.  We thus expect that the more complicated geometry of the GRMHD models will not obey a simple symmetry under field inversion. However, if the symmetric processes dominate over the antisymmetric processes (or vice versa), the $\vnet$ distributions could be approximately symmetric (or anti-symmetric) to an inversion of $\vb$. In section \ref{subsec:cp_distributions} we see that some models flip in $\vnet$ with inversion of $\vb$ suggesting that intrinsic CP emission ($j_V$) or Faraday rotation followed by Faraday conversion ($\rho_Q \rho_V$) are dominant.

\subsection{One-zone Model}\label{sec:one_zone}

We can crudely estimate the degree of circular polarization expected in Sgr A* and M87* using a one-zone model.  The model consists of a uniform sphere of hot plasma with radius $r = 5 GM/c^2$.  The uniform sphere has constant radiative transfer coefficients, the background radiation vanishes, and the spacetime is Minkowski.   

The magnetic field strength and number density of the plasma can be estimated using the one-zone model, following \citetalias{EHTC_2019_5,EHTC_2022_5} for M87* and Sgr A*.  As in those papers we assume the electrons have a thermal (Maxwell-J\"utner) distribution function.  The electron temperature and ion temperature need not be equal to each other as the plasma is collisionless and there may be  preferential heating of ions by turbulent dissipation \citep{quataert_heating_1999,yuan14,ressler_electron_2015,moscibrodzca16,zhdankin_twotemp_2021}.  We set the dimensionless electron temperature $\Theta_e \equiv k T_e/(m_e c^2) = 10$.  The magnetic pressure $B^2/(8\pi)$ is set equal to the gas pressure, assuming the ion-to-electron temperature ratio is $3$ and that the gas is pure hydrogen.  The angle between the magnetic field and line of sight is set to $60^\circ$.  The 1.3mm flux density is set equal to the observed $0.7$ Jy for M87* and $2.4$ Jy for Sgr A*.  Then the number density and magnetic field strength are
\begin{equation*}
	n_{e,\mathrm{M87*}} = 3.3 \times 10^4\;\mathrm{cm}^{-3};\  B_{\mathrm{M87*}} = 4.8 \; \mathrm{G}
\end{equation*} 
\begin{equation*}
	n_{e,\mathrm{Sgr~A*}} = 10^6\;\mathrm{cm}^{-3};\ B_{\mathrm{Sgr~A*}} = 29 \; \mathrm{G},
\end{equation*}
where we have iterated numerically over $n_e$ to find a solution.

Given the density and magnetic field strength we can compute the circular polarization fraction using the exact solution to the polarized radiative transfer equation \citep{deglinnocenti85, ipole18}).  We find 
\begin{equation}
	\mathrm{CP_{M87^*}} = 16.7\%
\end{equation}
and 
\begin{equation}
	\mathrm{CP_{SgrA^*}} = 5.1\%
\end{equation}
These CP values give an estimate of the resolved CP fractions ($\vavg$), which are consistent with per-pixel CP fractions in simulations (see Fig. \ref{fig:sample_ipole}, for example). In comparison, $\vnet$ observations (Table \ref{tab:cp_observations}) and $\vnet$ in simulations are much lower (Fig. \ref{fig:SgrA_distributions_MAD}, \ref{fig:M87_distributions_MAD}) which suggests cancellations across different parts of the image.

In the vicinity of the one-zone model parameters, we can probe the effects of intrinsic emission and Faraday conversion on Stokes $V$. The optical and Faraday depths ($\alpha_I r$, $\rho_Q r$ and $\rho_V r$) are approximately 0.4; in-between optically/Faraday thin and thick. In the optically/Faraday thin regime Stokes $V$ is well approximated by $j_V r$. In the moderately optically thick regime, Faraday effects are dominant; the solution to Stokes V with $j_V = \alpha_V = 0$ is similar to the full solution. At longer wavelengths, absorption effects play an important role. Further analysis of CP vs. frequency for the one-zone model is presented in appendix \ref{app:one_zone_cp_estimate}.

\section{Numerical Model}\label{sec:num_model}

\subsection{GRMHD Models}

We use a set of ideal GRMHD simulations for the KHARMA GRMHD simulation library in \citetalias{EHTC_2022_5} (hereafter "V3") and analyzed in detail in \citet{dhruv_inprep}. The models are run using KHARMA\footnote{KHARMA is a GPU-enabled, performance portable version of HARM \citep{gammie_harm_2003}.  It is publicly available at \url{https://github.com/AFD-Illinois/kharma}.}\citep{prather_grmhd_2022,prather_2024_kharma_zenodo} for $3\times10^4\; GM/c^3$.  GRMHD model parameters are described in detail in section \ref{subsec:LibraryParams}. The GRMHD models are nonradiative and therefore invariant under rescaling of the density of the plasma.  We choose a density scale (equivalently accretion rate or mass unit $\mathcal{M}$) so that the simulation flux density matches the observed flux density.

\subsection{Radiative Transfer Numerical Model}

We image the GRMHD simulation snapshots using the general relativistic ray-tracing code \ipole \citep{ipole18}.  The images are made by evaluating the intensity at a grid of points lying at the center of image pixels in a fictitious camera.  The photon trajectories are integrated backwards from the camera to, or past, the black hole.  Then the radiative transfer equation is integrated forward along the geodesic to the camera using the appropriate, relativistic version of Eq. \ref{eq:GenRadTrans} \citep{ipole18}.  

\subsection{Image Library Parameters}\label{subsec:LibraryParams}

The library parameters include both GRMHD model parameters and GRRT model parameters.  Our library has five parameters: two GRMHD and three GRRT:

\begin{enumerate}

	\item The magnetic flux through one hemisphere of the hole $\Phi_{BH}$, cast in dimensionless form $ \phi \equiv \Phi_{BH}(\dot{M} \left(GM/c^2\right)^2 c^{-1})^{-1/2}$. GRMHD models with $ \phi \sim 1 $ are known as Standard and Normal Evolution (SANE)  \citep{narayan12, sadowski_2013}. Models with $ \phi \sim 15 $ are know as Magnetically Arrested Discs (MAD) \citep{igumenshchev03, narayan03}.  MAD and SANE models are obtained by manipulating the magnetic field in the initial conditions.  
	
	\item Black hole spin $a_\star$, with $a_\star = $ -0.9375, -0.5, 0, 0.5, 0.9375. Negative spin indicates that the accretion flow is retrograde.
	
	\item The electron distribution function parameter  $\rhigh$ \citep{moscibrodzca16}, which sets the ion to electron temperature ratio $R = T_i/ T_e =R_{\mathrm{high}}\beta_{\mathrm{p}}^{2}/(1+\beta_{\mathrm{p}}^{2}) + 1/(1+\beta_{\mathrm{p}}^{2})$, where $\beta_\mathrm{p} $ is the ratio of the gas pressure to magnetic pressure. Typically $ \beta_{\mathrm{p}} $ is higher near the midplane than at the poles, so  high/low value of $\rhigh$ implies less/more emission contribution from the midplane. We set $\rhigh = $ 1, 10, 20, 40, 80 and 160 for M87*, and 1, 10, 40 and 160 for Sgr A*. 
	
	\item Inclination $\theta$ which is the angle between the wave vector and the orbital angular momentum of the accretion flow. The inclination is  $10^\circ$, $30^\circ$, $50^\circ$, $70^\circ$, $90^\circ$, $110^\circ$, $130^\circ$, $150^\circ$ and $170^\circ$ for Sgr A*. In models with $\theta<90$ ($\theta > 90$) the accretion disk rotates counter-clockwise (clockwise) on the sky. For M87* the black hole is imaged at an inclination of $17^{\circ}$ for a negative spin and $163^\circ$ for a non-negative spin, so that the inclination is chosen to match the large scale jet and the image asymmetry is chosen to match EHT images \citepalias[see][]{EHTC_2019_5}. 

	\item The sign of the magnetic field.  The field can be inverted without changing the GRMHD solution, so we have re-imaged all models with a reversed field.  We will use ``aligned field'' to refer to models with field near the poles that is parallel to the accretion flow orbital angular momentum, and ``reversed field'' to refer to models with polar fields that are antiparallel to the accretion flow orbital angular momentum.

	\end{enumerate}
	
Each model contains 600 images evenly spaced in the interval $15000 \mbox{--} 30000\; GM/c^3$ (1 $GM/c^3$ is $3 \times 10^5$ and $20$ seconds for M87* and Sgr~A* parameters respectively). The interval is chosen so that fluctuations associated with the initial conditions have damped away and the accretion rate is stable. The density scale $\mathcal{M}$ is fit every $5000\; GM/c^3$ to account for any potential depletion of mass in the accretion disk. For M87* and Sgr~A*, the average flux is within $5\%$ of $0.7$ Jy and $2.4$ Jy, respectively \citepalias[][\citealt{Wielgus2022lc}]{EHTC_2019_4}.

\section{CP for a Fiducial Model}\label{sec:cp_dist_one}

First consider a single, fiducial model: a MAD, spin +0.5, $\rhigh$ 160, inclination $30^\circ$ (and $150^\circ$) model for Sgr A*. This is one of the best-bet models based on EHT and multi-wavelength constraints \citepalias{EHTC_2022_5}.

\subsection{Sample Image}

\begin{figure*}
    \centering
    \includegraphics[width=0.8\textwidth]{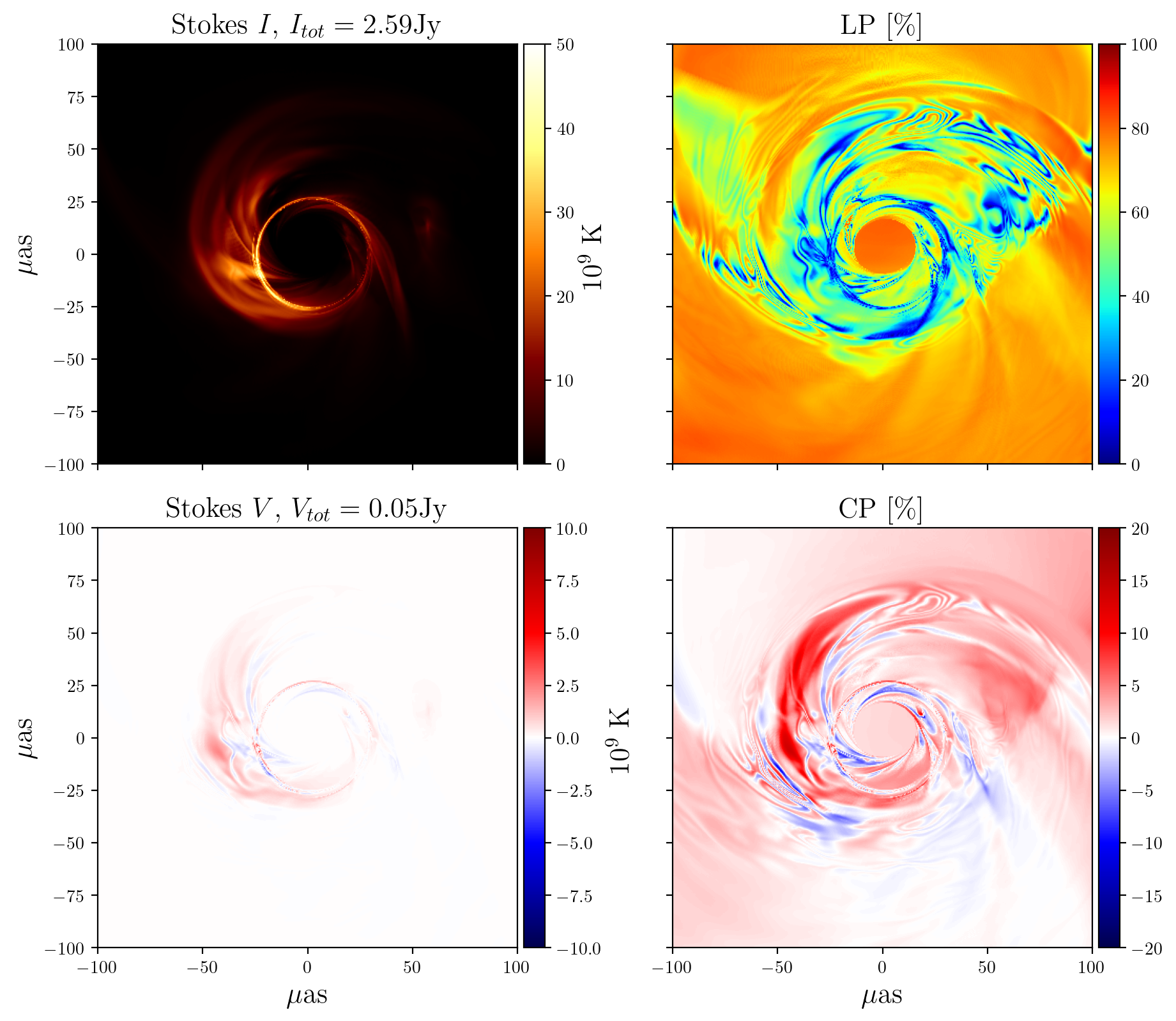}
    \caption{Sample MAD, $\textrm{spin} = +0.5$ at $2.75\times10^4 GM/c^3$, $\textrm{R}_\textrm{high} = 160$, inclination $30^\circ$, simulated Sgr~A* image using \texttt{ipole}. Top and bottom left is the Stokes I and V maps respectively in brightness temperature units. Top and bottom right panels is the fractional linear and circular polarization maps respectively. Resolution is 0.5$\mu$as per pixel.}
    \label{fig:sample_ipole}
\end{figure*}

Fig. \ref{fig:sample_ipole} shows a snapshot of Stokes I, Stokes V, LP fraction and CP fraction. One key feature of the image, typical of most of the models, is positive and negative fluctuations in CP that cancel out when $\vnet$ is evaluated.  In the image the CP fraction in individual pixels is as large as  10\textendash15\%, but integration over the image reduces the net CP fraction to 1\%. 

\subsection{CP Distribution}

\begin{figure}
    \centering
    \includegraphics[width=\linewidth]{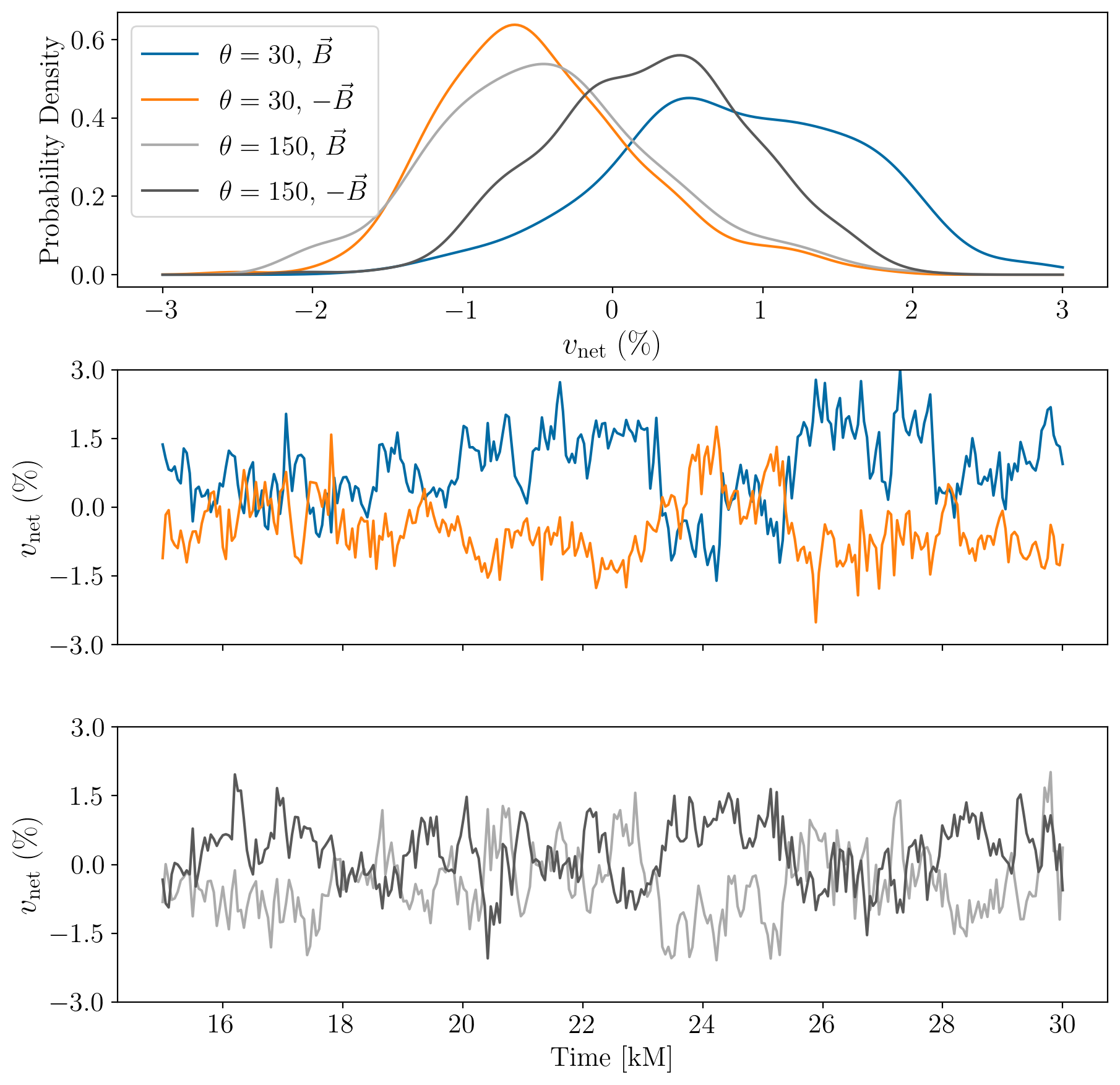}
    \caption{$\vnet$ for a Sgr~A* MAD, spin +0.5, $\rhigh$ 160, inclination $30^\circ$ (and $150^\circ$) model for both aligned and reversed field configurations. Top plot: Kernel density estimates of the distributions using kernel widths of 0.3\% to match observational errors in \citet{bower_alma_2018}. Bottom two plots: $\vnet$ lightcurves across 15000 $GM/c^3$ for each of these distributions with colors matching the legend above. Mean of the distributions parameterized by $(\theta,\pm \vb)$: $(30^\circ,\vb)=0.83\%$, $(30^\circ,-\vb)=-0.47\%$, $(150^\circ,\vb)=-0.42\%$, $(150^\circ,-\vb)=0.25\%$.}
    \label{fig:sample_distribution}
\end{figure}

The CP fluctuates in time.  To test a model we compare the model's distribution of CP to the observed distribution of CP.  The top panel in Fig. \ref{fig:sample_distribution} shows the distribution of CP broken down into the aligned and reversed field models, as well as the distribution seen from above (inclination $30^\circ$) and below (inclination $150^\circ$).  The time evolution is shown in the bottom two panels for each subset of the model.  Evidently reversing the field, or imaging from a complementary inclination, does not flip the distribution about 0, consistent with the discussion above; however there appears to be some anti-correlation with reversing the field which suggests intrinsic CP emission or Faraday rotation being important for $\vnet$ in these models. Notice that CP changes sign as a result of fluctuations in the small patches of polarization seen in Fig. \ref{fig:sample_ipole}.   

\subsection{Average Images}

\begin{figure*}
    \centering
    \includegraphics[width=0.9\textwidth]{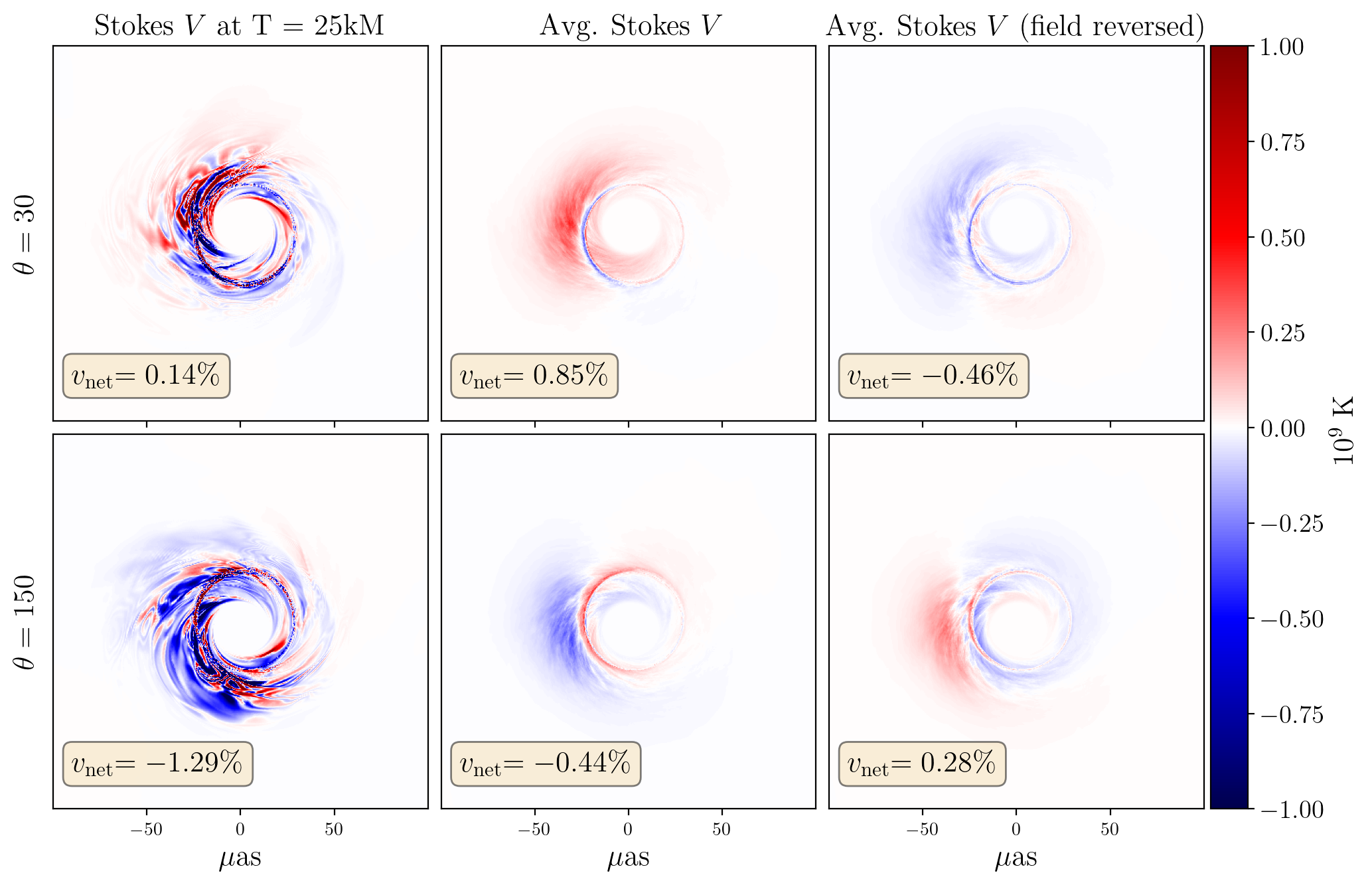}
    \caption{Snapshots and average images of Stokes V (in brightness temperature) for a Sgr~A* MAD, spin +0.5, $\rhigh$ 160, inclination $30^\circ$ (and $150^\circ$) model for both aligned and reversed field configurations.}
    \label{fig:sample_average_image}
\end{figure*}

Time-averaged images provide information about which CP generation mechanism dominates. The mean of the distribution of integrated CP {\em fraction} is not precisely equivalent to the integrated CP fraction of a time-averaged image, but if the total flux of each image is close to the mean value of 2.4 Jy then the two quantities are comparable. Fig. \ref{fig:sample_average_image} shows the average images for the fiducial model.

For this particular model, after averaging, the region that dominates the CP map is emission near the disk, close to the black hole \citepalias{EHTC_2019_5}. The bright positive feature is the region of the accretion disc where the fluid velocity is aligned with the line of sight and thus it appears prominently due to Doppler boosting. A clear ringlike structure is seen.  Its opposite sign is a consequence of the relatively low Faraday rotation thickness of the image and the imprint of the magnetic fields on Stokes V through Faraday conversion as observed in \citet{moscibrodzka_2021,ricarte_cp_2021}. The latter paper, in particular, shows that the sign of CP in the lensed photon ring always has the opposite sign of CP arising from Faraday conversion in Faraday thin images.

\subsection{Contribution of Transfer Coefficients}\label{subsec:coef_contribution}

\begin{figure*}
    \centering
    \includegraphics[width=0.95\linewidth]{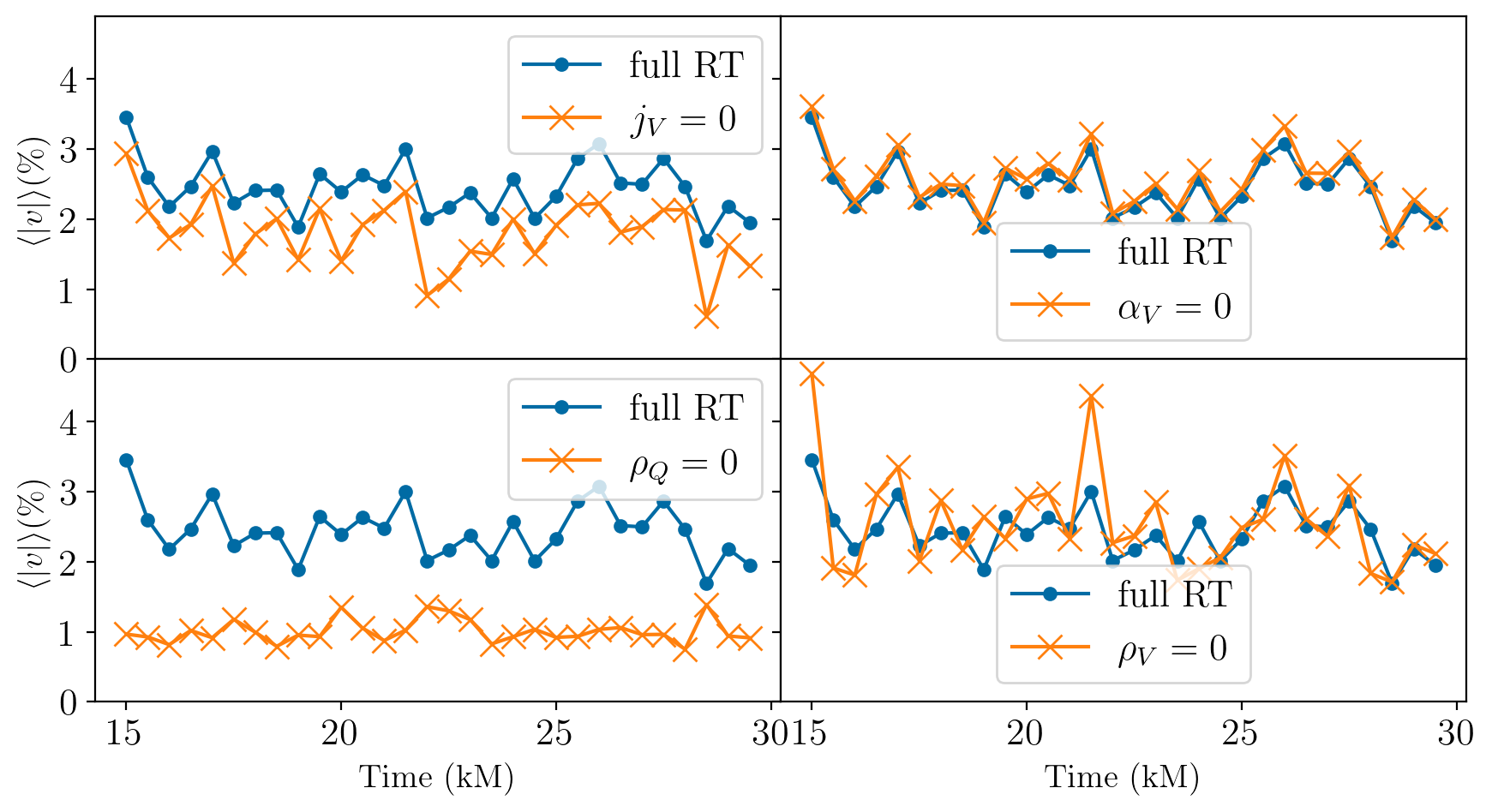}
    \caption{Comparison of $\vavg$ (resolved CP fraction) for a MAD spin +0.5, $\rhigh$ 160, inclination $30^\circ$ model across 15000M by re-imaging the model setting each coefficient to 0. The subplots each probe the effect of a process that influences CP, in clockwise order from top left: $j_V$ (intrinsic CP emission), $\alpha_V$ (CP absorption), $\rho_V$ (Faraday rotation) and $\rho_Q$ (Faraday conversion).}
    \label{fig:coef_compare}
\end{figure*}

Here we probe the relative importance of processes contributing to the net CP by turning off individual radiative transfer coefficients one by one.  We re-image  the fiducial model with 30 snapshots across 15000M, turning off $j_V$ (intrinsic CP emission), $\alpha_V$ (CP absorption), $\rho_Q$ (Faraday conversion) and $\rho_V$ (Faraday rotation).  Fig. 
\ref{fig:coef_compare} shows time series of CP in each case.

Evidently CP-selective absorption of unpolarized radiation ($\alpha_V$) plays a negligible role. This is because $\alpha_V$ is calculated from the Planck function and $j_V$ is about 2 orders of magnitude smaller than $j_I$ and the models are optically thin.  The remaining 3 coefficients all affect $\vavg$, with $\rho_Q$ the dominant mechanism of CP production, as $\vavg$ is highly suppressed when excluding Faraday conversion. Although $j_V$ is sub-dominant, it is non-negligible. This is also observed in MAD model images for M87\textsuperscript{*} at different $\rhigh$ values where we see the inclusion of $j_V$ can increase $\vavg$ by as much as 50\%. Faraday conversion and intrinsic CP emission both contribute to $\vavg$, but the effect of Faraday rotation varies. The importance of each coefficient also varies in time.  We conclude that only $\alpha_V$ is negligible but all remaining effects need to be accounted for to accurately model CP. 

A large number of models were investigated in a similar manner in \citetalias{EHTC_2023_9}, with one snapshot from each model that passed all polarimetric constraints, testing the relative effects of $j_V$, $\rho_V$ and $\rho_Q$. While the definitions of $\vavg$ in this paper, and $\langle| v |\rangle$ in \citetalias{EHTC_2023_9}, differ by a Gaussian blur of $20\mu$as in the latter, the results are consistent in that the contribution of $j_V$ is sub-dominant compared to $\rho_Q$.

The optical depths, magnetic field strength and thus CP production mechanisms vary greatly across models. Combined with the nontrivial effect of including the reversed field distributions, it is not possible to formulate a universally applicable, simple model for CP that is valid across all parameters.  By comparing the aligned and reversed field distributions, however, we can understand whether each model's CP is produced via $\vb$ field polarity invariant pathways (Faraday conversion through field twist) or non-invariant pathways (intrinsic emission or Faraday conversion through Faraday rotation).

\section{CP Distributions Across All Models}\label{subsec:cp_distributions}

\begin{figure*}
    \centering
    \includegraphics[width=\textwidth]{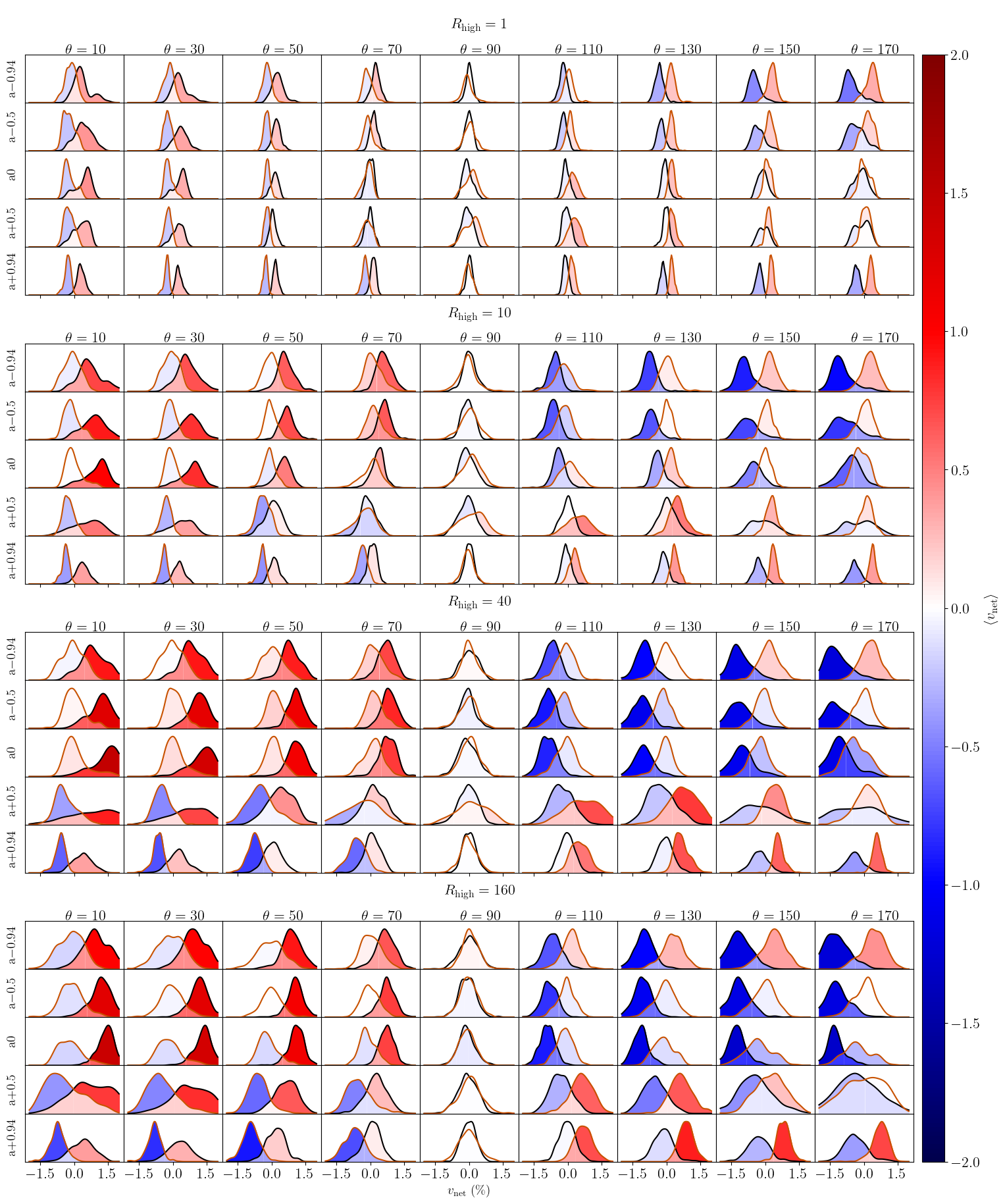}
    \caption{Distributions of $\vnet$ for all the MAD Sgr~A* models. For each subplot, the x-axis corresponds to $\vnet$ and the y-axis corresponds to the Probability Density Function (PDF) of the model. Across subplots, the Y-axis spans spin, grouped in $\rhigh$. The X-axis spans in observer inclination. The black (orange) lines represent the aligned (reversed) field distribution respectively. The color filled within the distributions corresponds to the mean $\vnet$ and the color in the overlapping region shows mean $\vnet$ of both field configurations combined. The height of each subplot is adjusted so that the maximum of the distribution has the same height in all panels.}
    \label{fig:SgrA_distributions_MAD}
\end{figure*}

\begin{figure*}
    \centering
    \includegraphics[width=\textwidth]{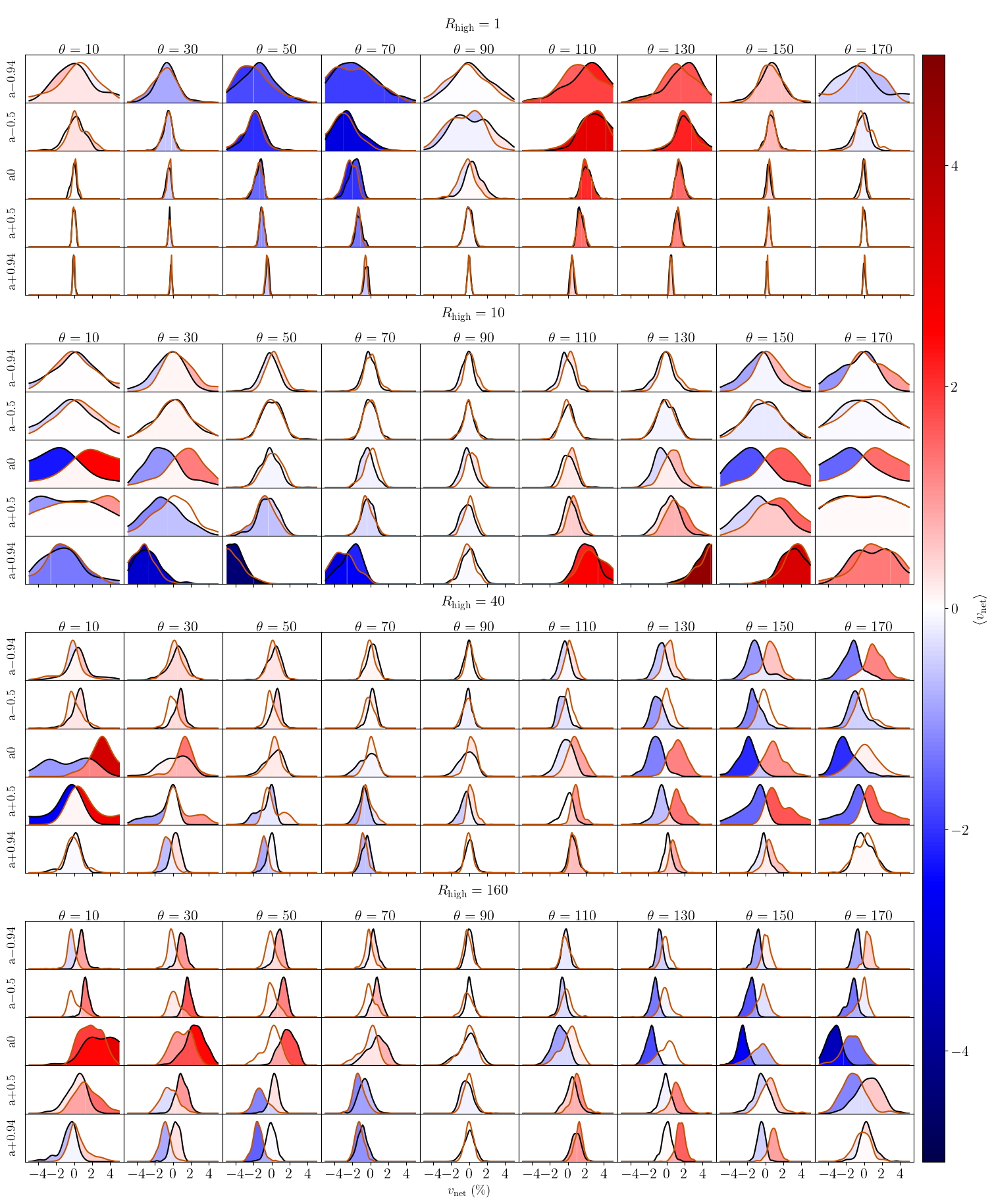}
    \caption{Same as Fig. \ref{fig:SgrA_distributions_MAD} except for SANE distributions.}
    \label{fig:SgrA_distributions_SANE}
\end{figure*}

Convergence tests of distributions of $\vnet$ are given in appendix \ref{app:cp_convergence} -- the $\vnet$ distributions appear converged with respect to GRMHD resolution and GRRT image resolution. Fig. \ref{fig:SgrA_distributions_MAD}\textendash \ref{fig:SgrA_distributions_SANE} show the $\vnet$ distributions (aligned and reversed field) of Sgr~A* for all models in the library.  M87* distributions are given in appendix \ref{app:m87_vnet}. First and second moments of each distribution are given in appendix \ref{appsec:cp_tables}.

Comparing MAD and SANE models, we see that SANE models have higher $\vnet$, particularly for lower $\rhigh$ values. For MAD models almost all models have $|\vnet|<2\%$ whereas SANE models have snapshots with $|\vnet|>4\%$. Given the low detected values of CP for Sgr~A* and M87*, some of our SANE models can be ruled out. MAD models also exhibit cleaner trends across model parameters whereas SANE (especially low $\rhigh$) models are more turbulent. \citet{ricarte_cp_2021} (see Fig. 8, 9, 13) also find that SANE models have higher Faraday depths than MADs. Higher Faraday depth implies more scrambling of linear polarization, and thus, to the extent that Faraday conversion is important, scrambling of circular polarization. This scrambling hides imprints of the magnetic field in CP measurements of SANEs compared to MADs. 

The effect of field reversal is mixed. For some SANE models, $\vnet$ is nearly antisymmetric (spin 0, $\rhigh$=10) while a few models $\vnet$ is nearly symmetric (spin 0, $\rhigh$=1).  A majority of the SANE models, and all of the MAD models, show imperfect symmetry/antisymmetry under field reversal, indicating that both magnetic field twist and Faraday rotation + intrinsic emission contribute significantly.
    
In Sgr~A* CP is almost 0 for edge-on models. This can be attributed to the cancellation that occurs across every image due to symmetries in the magnetic field geometry (see \citealt{tsunetoe_polarization_2021, ricarte_cp_2021} for a detailed description). Edge-on models tend to have higher Faraday depths, which also contributes to increased cancellation of CP across the image. We find that Faraday rotation depths for SANE models are two orders of magnitude higher than corresponding MAD models. Faraday depth is a strong function of $\rhigh$ (increases), spin (decreases from retrograde to prograde) and inclination (increases till $90^\circ$). $\rhigh$ and spin directly influence the temperature of the electrons and models with hotter electrons have lower Faraday depths.

\subsection{Parameter Dependence of CP Distributions}\label{subsec:cp_dist_prop}

Here we focus on parameters that can influence $\vnet$ across all models such as spin, inclination $\rhigh$ and frequency.

\subsubsection{Spin Dependence}

Black hole spin influences the sign and shape of the $\vnet$ distributions. Prograde spin model snapshots contain components with opposite signs of $\vnet$, i.e., more spatial cancellation than in low spin models. As a result, high spin prograde model distributions of $\vnet$ are broader than low spin models.

An image in Stokes I (or V) can be divided into the weakly lensed component (n=0) and a strongly lensed component where photons wrap around the black hole in n half circles (n=1,2,3...) \citep[see][]{johnson_photon_2020}. Each ring $n$ is $\sim \exp(-\pi)$ fainter than the next.  In this work we see effects primarily from the n=0 and n=1 images. 

\begin{figure*}[h]
    \centering
    \includegraphics[width=\textwidth]{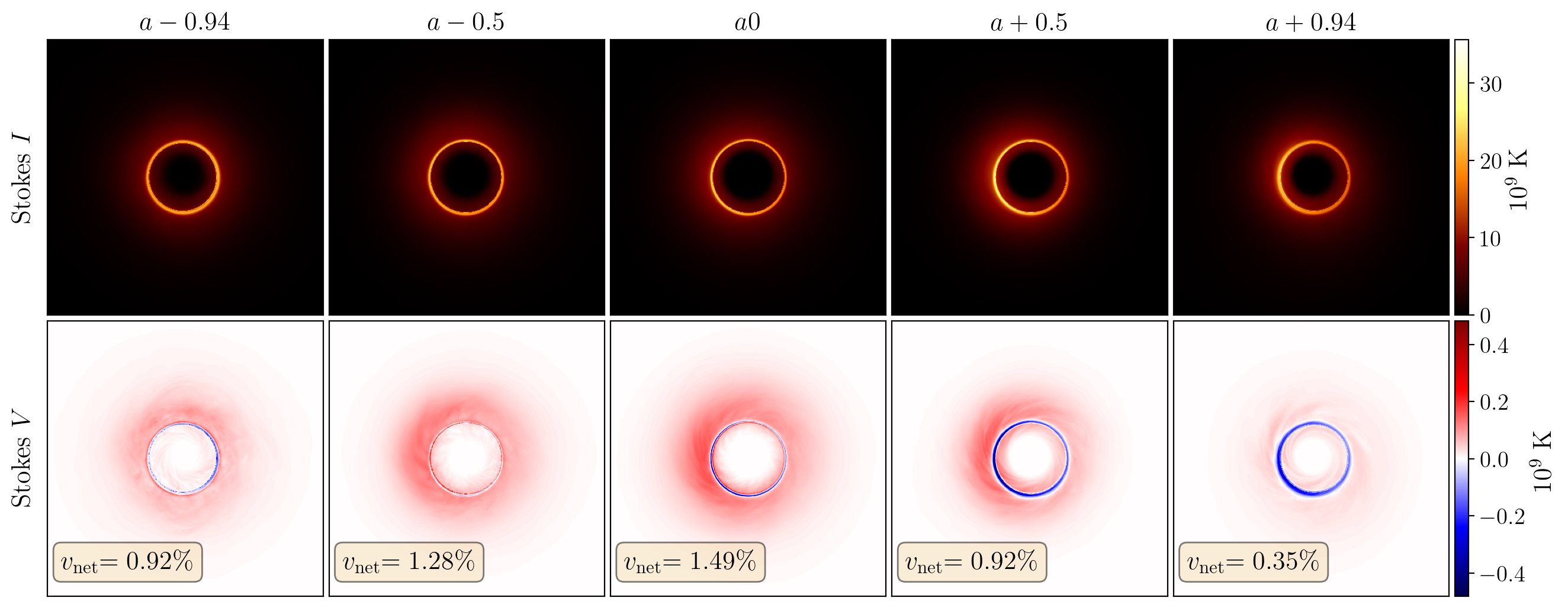}
    \caption{Average images of Stokes I and V across spin for Sgr~A* MAD models at $\rhigh$ 40, inclination $10^\circ$.}
    \label{fig:MAD_spin_average}
\end{figure*}

For face-on prograde MAD models, the opposite signed n=0 and n=1 portions of the image becomes important, with the n=1 ring becoming the dominant source of fractional CP as seen in Fig. \ref{fig:MAD_spin_average}, which show average images of CP for a MAD model across spin. The n=0 mode of the CP image has contributions from intrinsic emission and Faraday conversion both through Faraday rotation and twist. In the Faraday thin and optically thin regions with toroidal magnetic fields, the n=1 photon ring is the opposite sign of the n=0 component. The opposite sign of the photon ring is a consequence of Faraday conversion through the twist of the magnetic field as explored in \citet{ricarte_cp_2021}. The field structure of retrograde models is less toroidal and thus the n=1 contribution is reduced. This effect is also reduced as Faraday thickness increases and thus is less prominent in retrograde and SANE models. While the sign flip in the n=1 photon ring is prominent at face-on inclinations, this effect of prograde models having opposite signed $\vnet$ distributions is also seen at higher inclinations although it is uncertain if the same phenomenon is responsible. It is possible that as spin increases the contribution of n=0 portion of the image decreases, however further investigation would be necessary.

\subsubsection{Inclination Dependence}

Images at observer inclination $\theta \neq 90^\circ$ typically contain contributions from the near and far-side regions ($z>0$ and $z<0$ where $z=0$ is the midplane of the system). For both of these regions $\vec{k} \cdot \vb$ is important, along with the direction of twist of $\vb$ along the geodesic. At low inclinations, photons from the far-side region have larger optical and Faraday depths from gravitational lensing increasing the path length. For higher $\rhigh$ models, the cool disk may also increase the Faraday depth for far-side photons travelling through it. As a result, the far-side contribution can be scrambled or even change sign in certain regions, but it is unlikely to exactly cancel out, or surpass the near-side component thus generating a net $\vnet$ biased towards the near-field component.

For inclinations $\theta<90^\circ$ and aligned magnetic fields, MAD models on average have positive $\vnet$ (ignoring photon ring effects). In the near-side region, $\vec{k}\cdot\vb>0$ so $j_V>0$ and $\rho_V>0$. For Faraday thin regions, $\rho_V>0$ increases Stokes U which in turn increases Stokes V via conversion (see Eq. \ref{eq:GenRadTrans}). Faraday conversion through the twist in magnetic fields also contributes towards $\vnet>0$: \citet{ricarte_cp_2021} demonstrate this for a face-on model by considering the twist of $\vb$ along the jet as it (the jet) broadens out (denoted as the vertical twist $\xi_V$). The vertical twist $\xi_V$ (and the Stokes basis rotation) is clockwise along the line of sight which corresponds to a rotation of $Q>0$ to $U>0$ and thus $V>0$ as $\rho_Q>0$. Another form of twist explored in \citet{enslin_2003,ricarte_cp_2021}, the transverse twist ($\xi_T$), occurs for edge-on models. For disks rotating counter-clockwise in the sky ($\theta<90^\circ$), trailing magnetic fields embedded in the disk will be twisted counter-clockwise along the line of sight (for the approaching jet), leading to $V<0$. As this effect is maximal for edge-on simulations, its imprint on inclination compared to that of the vertical twist is minimal.

While it is difficult to ascertain the sense of twist of $\vb$ for intermediate observer inclinations, one can expect the effect of twist along the jet to reduce as the observer inclination increases. Due to symmetries in the global magnetic field across the midplane, as $\theta \rightarrow 90^\circ$, any additional effects of twist should cancel out. Also, a complementary inclination angle will reverse $\vec{k}\cdot \vb$ and twist of $\vb$, thus reversing the sign of CP. Thus, on average, the $\vnet$ distributions should follow a cosine function but fluctuations will prevent individual snapshots from following a neat pattern.

From Figs. \ref{fig:SgrA_distributions_MAD} and \ref{fig:SgrA_distributions_SANE}, we see that this is the case for the aligned and reversed field models. The universality of this behavior across all models suggests that the mean across field configurations encodes the sense of twist of the magnetic field; since CP generated from conversion of LP via a twisted magnetic field is the only mechanism invariant to the sign of $\vb$. 

A mean positive $\vnet$ thus implies an overall clockwise twist in $\vb$ along $\vec{k}$ for both retrograde and nonspinning models with $\theta < 90^\circ$ (where the $n=0$ contribution is dominant). It could also imply an overall counter-clockwise twist for prograde models with $\theta>90^\circ$, only in cases where the $n=1$ contribution is greater than the $n=0$ component which is most $a+0.94$ models and a few $a+0.5$ models.

A mean negative $\vnet$ implies the same but with the corresponding opposite sense of twist and observer inclinations. Thus, we find that the sign of the mean $\vnet$ is sensitive to not only the global sign of the magnetic field, but also the sense of rotation of the accretion flow.

\subsubsection{$\rhigh$ Dependence}

As mentioned in section \ref{subsec:LibraryParams}, higher $\rhigh$ implies cooler electrons in the disk. This causes less emission from the midplane of the disk. For higher $\rhigh$, a higher mass unit ($\mathcal{M}$) is required to obtain the same output flux as that from a lower $\rhigh$ value simulation, and this implies higher density and therefore an increase in all the radiative transfer coefficients. This causes two important effects. One is an increase in circular polarization for individual pixels because of increased Faraday conversion and emission. The other is a decrease in overall circular polarization arising from an increase in depolarization: scrambled LP from Faraday rotation leads to scrambled CP through Faraday conversion.  The presence of a cooler, denser population of electrons in the disk midplane is particularly important for increasing Faraday rotation.

\begin{figure}
    \centering
    \includegraphics[width=0.95\linewidth]{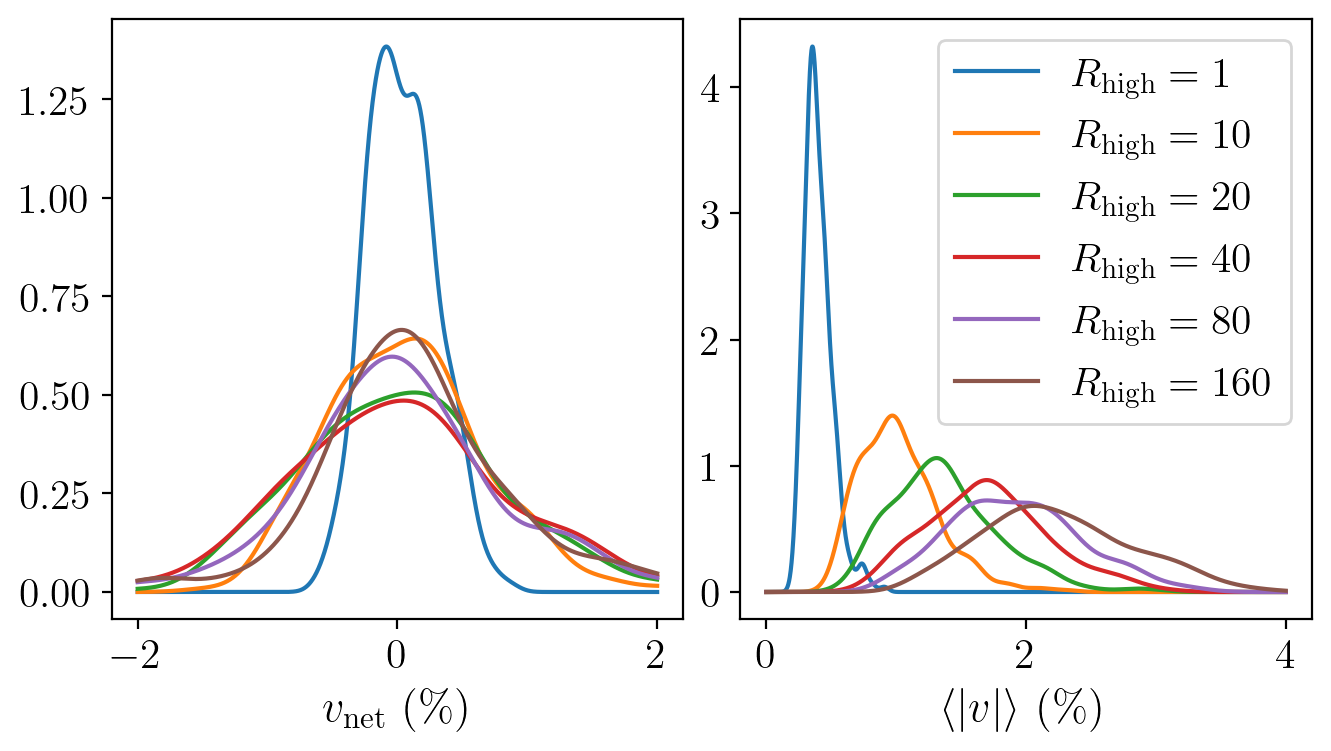}
    \caption{Kernel density estimations of the $\vnet$ and $\vavg$ distributions for a MAD, spin +0.5, M87 black hole for different $\rhigh$ values. While there is no observable trend for $\vnet$, there is a clear increase in $\vavg$ with $\rhigh$.}
    \label{fig:CP_Rhigh}
\end{figure}

The overall effect is that $\vnet$ (unresolved CP) distributions broaden with increasing $\rhigh$, and $\vavg$ (resolved CP) distributions increasing in magnitude with $\rhigh$. The distributions of $\vnet$ and $\vavg$ for different values of $\rhigh$ and M87 parameters is given in Fig. \ref{fig:CP_Rhigh}. Although it is difficult to observe any trend in $\vnet$ (likely due to cancellations), a clear increase in $\vavg$ with $\rhigh$ is seen. As this effect is seen for most spins, we conclude that resolved observations of Stokes V ($\vavg$) in the future can constrain electron temperature models.  This is consistent with results found in \citet{moscibrodzka_2021}.

 Models with $R_{\mathrm{high}}=1$ are qualitatively different in $\vnet$ especially in the SANE models, as these models have hot, dense discs which dominate the emission in a relatively concentrated region (in the poloidal direction) in the midplane. Higher $\rhigh$ models in comparison require a higher scaling factor to obtain the same flux. This causes more contributions from both near and far jet sheath regions, which combined with a cooler midplane can increase the Faraday depths and emission regions.

\subsubsection{Frequency Dependence}

\begin{figure*}
    \centering
    \includegraphics[width=\textwidth]{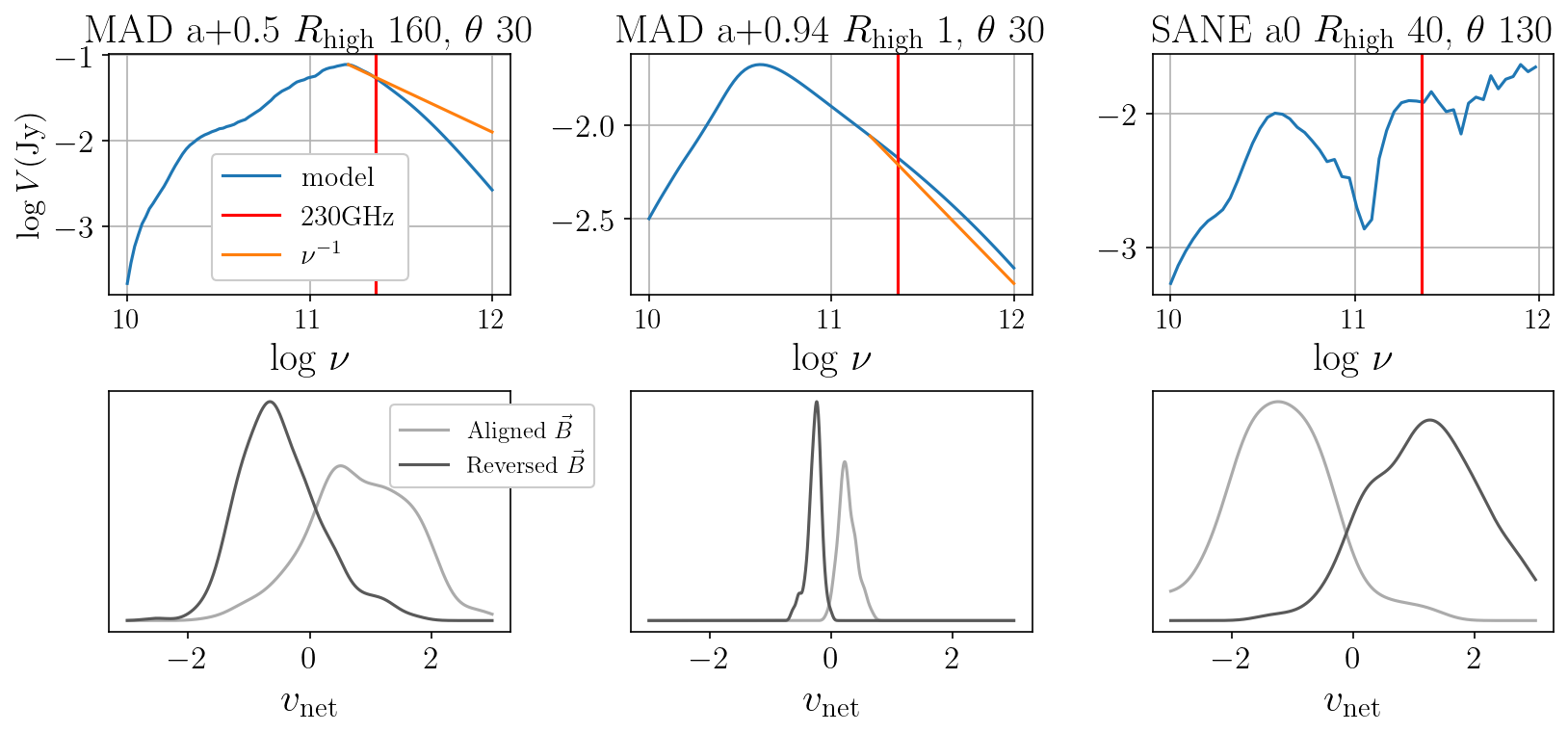}
    \caption{Upper row: image-integrated Stokes $V$ for three different GRMHD models across frequency. $\nu^{-1}$ is plotted for comparison in the Faraday thin regime. Red line indicates 230GHz (1.3mm), the frequency at which the model library is generated. Bottom row: distributions of $\vnet$ in time for both field configurations of the corresponding models at 230GHz. The behavior of $\vnet$ with an inversion of field can inform the dominant CP mechanism which affects the spectral behavior.}
    \label{fig:stokesv_freq}
\end{figure*}

While not a parameter explored in the image library, we investigate $\vnet$ of a few models versus frequency, shown in Fig. \ref{fig:stokesv_freq} for three models: (MAD a+0.5, $\rhigh$ 160, inclination $30^\circ$), (MAD a+0.94, $\rhigh$ 1, inclination $30^\circ$) and (SANE, a0, $\rhigh$ 40, inclination $130^\circ$). We find three different spectral behaviors of Stokes $V$, which suggests different mechanisms are dominant in each model. The SANE model, while optically thin, remains Faraday thick. Thus even at higher frequencies when probing inner regions of the accretion flow, the Stokes $V$ signal does not significantly decrease. The MAD models are Faraday thin near 230GHz so Stokes $V$ reduces as frequency increases and the $\rhigh=160$ model decreases faster compared to the $\rhigh=1$ model. From distributions of $\vnet$, we find that the MAD $\rhigh$ 160 model does not neatly change sign with $\vb$ reversal whereas the $\rhigh$ 40 model does. This suggests that Faraday conversion (through twisted magnetic fields) is important to the former (high $\rhigh$ model) but not the latter (lower $\rhigh$) where intrinsic emission or Faraday conversion through rotation dominate.

In appendix \ref{app:one_zone_cp_estimate}, we analyze the analytic solution to Stokes $V$ for a single geodesic and find that Stokes $V$ from just Faraday conversion decreases much faster with increasing frequency than Stokes $V$ from intrinsic emission. Qualitatively, comparing these analytic results (Fig. \ref{fig:stokesv_analytic_freq}) to the numerical models (Fig. \ref{fig:stokesv_freq}) suggests that a steep $\nu^{-2}$ scaling in Stokes $V$ as seen in the MAD a+0.5 $\rhigh$ 160 model suggests a Faraday conversion dominated model and a slow $\nu^{-1}$ scaling as seen in the MAD a+0.94 $\rhigh$ 1 model is intrinsic emission dominated. The frequency scaling estimates are heuristic and an extensive frequency analysis across all models is needed for direct comparison with observations.

\subsection{Fits to $\vnet$ distributions}

The mean and standard deviations of all $\vnet$ distributions are provided in Appendix \ref{appsec:cp_tables}.  Likely because of the mix of physical processes that contributes to CP it is difficult to provide a simple fit to these moments that covers the entire model space.  The Sgr A* MAD models, however, exhibit clear trends.  The mean is readily fit by  
\begin{equation}\label{eq:sgra_mad_mean_fit}
    \langle \vnet\rangle = 0.61\% \, \cos(\theta)\left(1 - \frac{0.64}{\textrm{R}_\textrm{high}}\right)\left(S - \frac{(1+a)^2}{3}+1\right)
\end{equation}
which we extracted using the  \texttt{PySR} symbolic regression code \citep[][]{pysr,cranmer2020discovering} .  Here $\theta$ is the observer inclination and $S$ is the sense of the magnetic field ($1$ for aligned and $-1$ for reversed cases). The fitting function recovers a cosine dependence on the observer inclination, which we attribute to the overall switch in sign from viewing at opposite poles, and increasing cancellations when observing edge-on due to symmetries of the magnetic field.

\begin{figure}
    \centering
    \includegraphics[width=\linewidth]{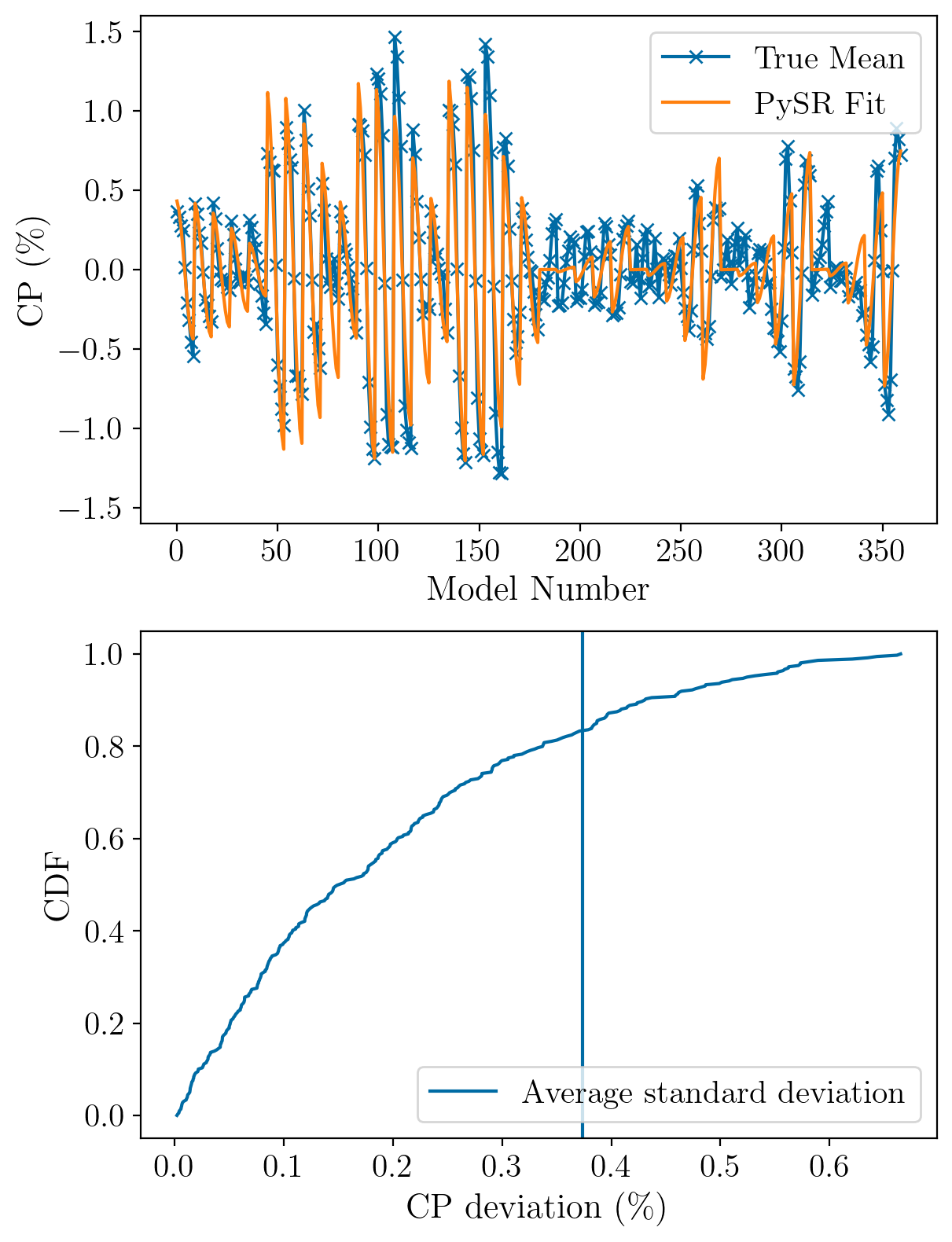}
    \caption{Top: Performance of \texttt{PySR} fitting function inferences for the means of Sgr~A* MAD CP distributions. Model numbers span in the order of inclination, spin, $\rhigh$ and field configuration. Model numbers 0-180 correspond to aligned field models with 181-360 being reversed. Bottom: The cumulative distribution function (CDF) of the L1 norm of error for the fitting function. Solid line is the average standard deviation of CP in the models.}
    \label{fig:pysrfit_sgra_mad_mean}
\end{figure}

The mean $\vnet$ and the fit (Eq. \ref{eq:sgra_mad_mean_fit}), are shown in Fig. \ref{fig:pysrfit_sgra_mad_mean}. The fit accurately measures the mean $\vnet$ of roughly 80\% of the models to within 0.3\% (absolute difference). The fits for the mean and standard deviations of SANE models do not permit such an accurate fit, but the raw data is provided in appendix \ref{appsec:cp_tables}.

\section{Comparison to Observational Data}\label{sec:constraints}

\subsection{Sgr~A*}

\begin{figure*}
    \centering
    \includegraphics[width=0.9\linewidth]{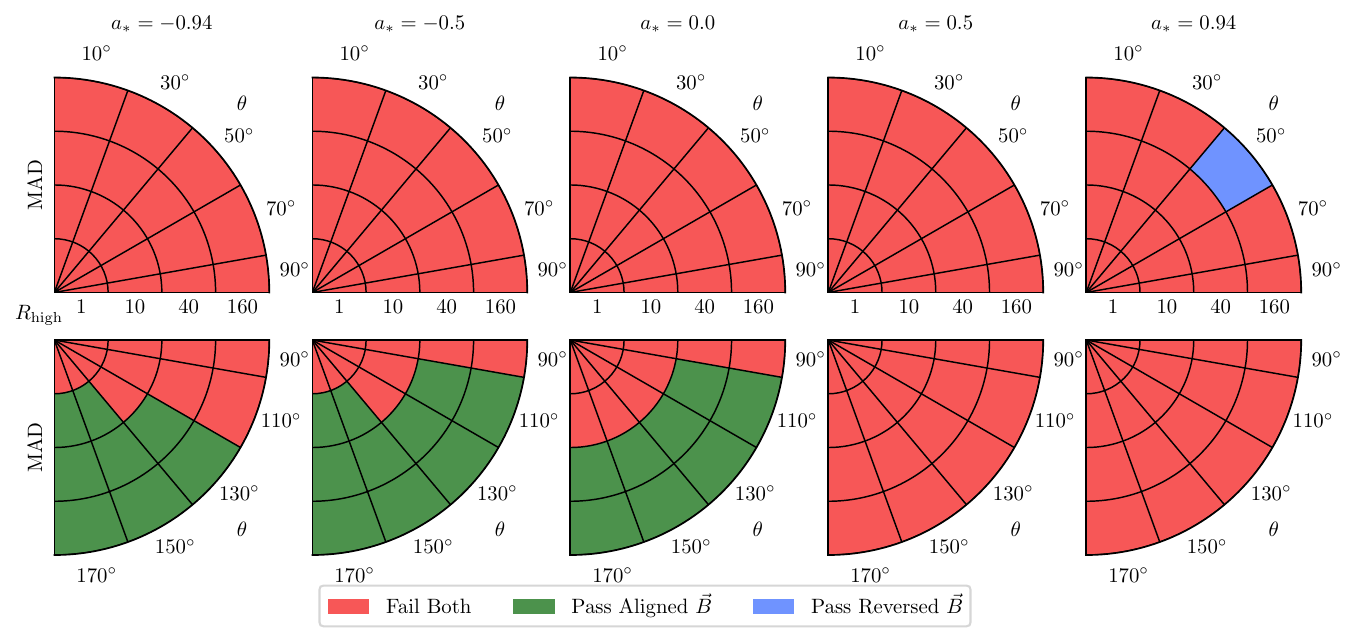}
    \centering
    \includegraphics[width=0.9\linewidth]{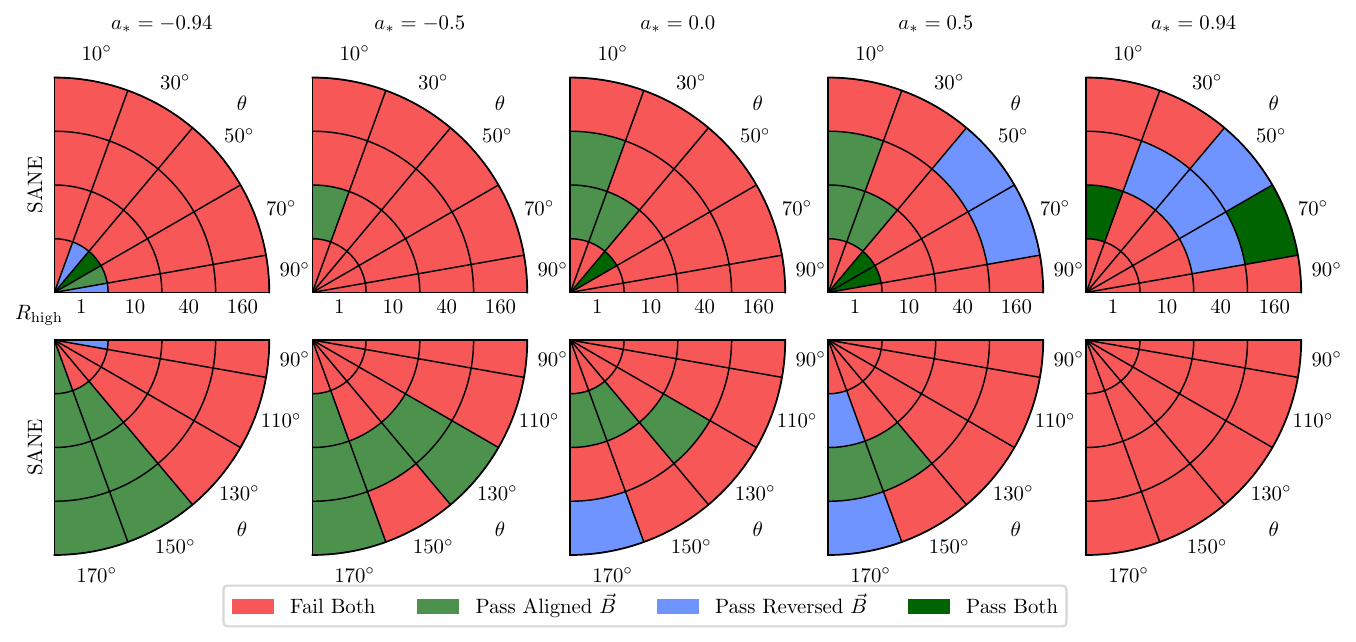}
    \caption{Pass/fail plots of the Illinois Sgr~A* model library for $\vnet$ (top row MAD models, bottom row SANE models). Each pie plot represents a given spin, with $R_{\rm high}$ along the radial direction and inclination ($\theta$) as the polar angle of the subplot.}
    \label{fig:constr}
\end{figure*}

Which models produce CP that matches Sgr A* and M87*? Detections and limits of $\vnet$ for M87* and Sgr~A* are given in table \ref{tab:cp_observations}. For Sgr~A* the observations from ALMA given in \cite{bower_alma_2018} and \cite{wielgus_orbital_2022} are used as a constraint on the CP (8 data points, assumed uncorrelated, ranging from $-1.5\%$ to $-0.92\%$). To test the models, a procedure similar to that in \citepalias{EHTC_2022_5} is used. Two-sided Kolmogorov-Smirnov (KS) test p-values ($p$) are computed between distributions of the model and observations. The model is sampled every 400M to obtain an approximately uncorrelated sample. The model fails the constraint if both the aligned and reversed field distribution gives $p<0.01$, giving 99\% confidence in rejecting the null hypothesis that the model and observations arose from the same distribution. Constraint plots are shown in Fig. \ref{fig:constr}. 

\setlength\tabcolsep{0pt}
\begin{table}
    \centering
    \caption{Measurements of CP fraction at 1.3mm for EHT targets. Note that for \citet{wielgus_orbital_2022}, the epoch values are averages of 3 hour windows. While we report epoch values for \cite{munoz_circular_2012}, we only use the ALMA epoch $\vnet$ for comparing our simulations with Sgr~A*.}
    \begin{tabular*}{0.99\linewidth}{@{\extracolsep{\fill}}cccc}
         Reference & Source &\begin{tabular}{c} Mean \\ $\vnet(\%)$\end{tabular} &\begin{tabular}{c} Epoch \\ $\vnet(\%)$\end{tabular} \\
         \hline
         \citet{munoz_circular_2012} & Sgr~A* & $-1.2 \pm 0.3$ & \begin{tabular}{c}
              $-1.1$\\
              $-1.2$\\
              $-1.1$\\
         \end{tabular}\\
         \hline
         \citet{bower_alma_2018} & Sgr~A* & $-1.1 \pm 0.2$ & \begin{tabular}{c}
              $-1.3$\\
              $-0.9$\\
              $-1.3$\\
         \end{tabular}\\
         \hline
         \citet{Goddi2021} & Sgr~A* & $[-1.0, -1.5] \pm 0.6$ & N/A \\
         \hline
         \citet{Goddi2021} & M87* & $\lesssim \pm 0.8$ & N/A \\
         \hline
         \citet{wielgus_orbital_2022} &\, Sgr~A* & $-1.23 \pm 0.4$ & \begin{tabular}{c}
              $-1.5$\\
              $-1.4$\\
              $-0.9$\\
              $-1.0$\\
              $-1.0$\\
         \end{tabular}\\
         \hline
    \end{tabular*}
    
    \label{tab:cp_observations}
\end{table}

We have $27/180 = 15\%$ of the MAD models pass and $43/180 = 24\%$ of the SANE models pass the CP constraint. SANE models are less constrained overall than MAD models. This can be attributed to higher fractional CP generated in general increasing the chance that one combination of flow orientation and field configuration produces sufficiently negative CP to pass the constraint.

All the best-bet models given in \citepalias{EHTC_2022_5} fail the CP constraint. These  are prograde MAD models with high $\rhigh$ at a low inclination. This can be attributed to the effect of black hole spin on prograde models as mentioned in section \ref{subsec:cp_distributions} \textemdash cancellation between the n=1 photon ring and n=0 "weakly lensed" emission shifts the distributions towards the center while the observations are predominantly around the -1\% level.

Given that $\theta>90^\circ$ represents a clockwise rotation of the disc in the sky (beyond the ergosphere for retrograde models), there is a clear preference for retrograde MAD models in which the flow orientation is clockwise in the sky. All these models pass when the $\vb$ field is oriented parallel to the disk angular momentum vector (aligned field). This is because the photon ring has a negligible effect and the overall n=0 emission (in this configuration) is negative. These findings are consistent with the GRAVITY measurements of the orientation of flow for Sgr~A* based on the motion of NIR flares \citep{gravity_clockwise_2018}.
    
For the one passing prograde MAD model, the field is in the reversed configuration meaning that all the passing MAD models have the dipole moment of the field pointed away from the observer. Given the combination of inclinations and field orientations, we find that the $\vnet$ constraint (on MADs) is sensitive to both the sense of rotation of the flow \citep[as proposed and explored in][respectively]{enslin_2003,moscibrodzka_2021} and the overall direction or structure of the magnetic field configuration \citep{beckert_falcke_2002}. If the constraint were insensitive to the sense of rotation of the flow, then models with $\theta<90^\circ$ would pass and if it were insensitive to the direction of magnetic field, then both aligned and reversed fields for all passing MAD models would pass.

All the MAD edge-on inclination models fail and only one of the SANE edge-on inclination models pass the CP constraint. This is because of cancellations that occur across the image domain due to symmetries in the magnetic field structure for each of the models (\citet{ricarte_cp_2021}), yielding a net zero CP fraction.


Combined with the constraints in \citepalias{EHTC_2022_5}, the CP test eliminates all models. The best-bet prograde MAD models do not produce sufficient CP whereas the retrograde MADs (which pass CP constraints) are mostly eliminated from the m-ring constraints, which compare the ring width, asymmetry and diameter of the model to the observed image. Most of the SANEs fail m-ring and non-EHT (multi-wavelength) constraints.

\subsection{M87*}

CP measurements of M87* given in \citet{Goddi2021} constrain $|\vnet|<0.8\%$. This constraint was used for model comparison in \citepalias{EHTC_2021_8} and most models aside from a few SANE models contained snapshots with $\vnet$ within this value. While the models used in this paper are from different GRMHD models evolved out to longer timescales, the underlying physics remains the same. Given the frequent $\vnet$ sign fluctuations observed across all models, there exist a few snapshots where $|\vnet|<0.8\%$. A few SANE spin 0 models have only $2-10\%$ of snapshots passing this constraint but most of the models have many passing snapshots as the distributions are close to 0. The discerning power of $\vnet$ for M87*, while broadly consistent with the results of \citetalias{EHTC_2021_8} do not reveal any significant trend.

While we do not apply the resolved circular polarization upper limit of 3.7\% as given in \citetalias{EHTC_2023_9}, we do not expect our results to be significantly different, as the GRMHD libraries contain the same underlying physics and differ mostly in the final integration time. \citetalias{EHTC_2023_9} utilizes libraries of the reversed field configuration and properties of the polarimetric quantities do not appear to vary significantly aside from $\vnet$ and the angle of the axisymmetric Fourier component of the EVPA ($\angle \beta_2$ in the paper), which is consistent with expectations. Incorporating the additional parameter of the field configuration to the library, the outcome is not significantly changed, with MAD models with R\textsubscript{low} 10 models still preferred.

\section{Discussion}\label{sec:discussion}

Here we discuss certain limitations of the theoretical models and provide some expectations for future improvements. As these models are the same underlying GRMHD simulations used in the \kharma library in \citepalias{EHTC_2022_5}, most of the caveats discussed in that paper apply here. Given the diversity of the model parameters, it is difficult to predict the effect of an improvement on an entire library of simulations and thus each of the suggestions merits a separate analysis that is beyond the scope of the paper.

The primary caveat is that the models are highly variable in Stokes I and V when compared to Sgr~A*. While $\vnet$ for Sgr~A* is consistently observed at the percent level, frequent sign crossings are observed for most models due to turbulence and rapidly changing optical and Faraday depths. Comparing the mean of distributions could provide a more robust method of comparing simulations to observations (\citetalias{EHTC_2022_5},\citet{Wielgus2022lc,Wielgus2023}). However for our models the sign crossings invariably produce mean $\vnet$ values lower than the percent level. Improvements to physics in GRMHD models such as including self-consistent electron heating and cooling mechanisms or addition of leading order collisionless corrections such as viscosity and heat conduction as given in \citet{chandra_extended_2015} can potentially reduce the fluctuations of Stokes V in the models and thus shift the $\vnet$ distributions away from 0.

The initial condition of the current GRMHD library is an equilibrium Fishbone-Moncrief torus solution \citep{fishbone_relativistic_1976} seeded with a magnetic field. Most of the emission regions for such simulations occur within $20M$ with the time period chosen so as to allow these regions to reach a steady state solution. However, alternative initial conditions such as stellar-wind-fed models \citep{ressler_ab_2020} can yield qualitatively different results in all the Stokes images. Magnetic fields can greatly influence the structure of the CP image. Since the magnetization of the stellar winds is poorly constrained, qualitative features of the polarimetric image are sensitive to the choice of plasma $\beta$ and further studies are necessary. For higher magnetizations of the stellar wind, the simulations can be expected to settle into a MAD state and show similar properties of the corresponding MAD simulation with similar inclinations and electron temperatures in our library.

\begin{figure}
    \centering
    \includegraphics[width=0.9\linewidth]{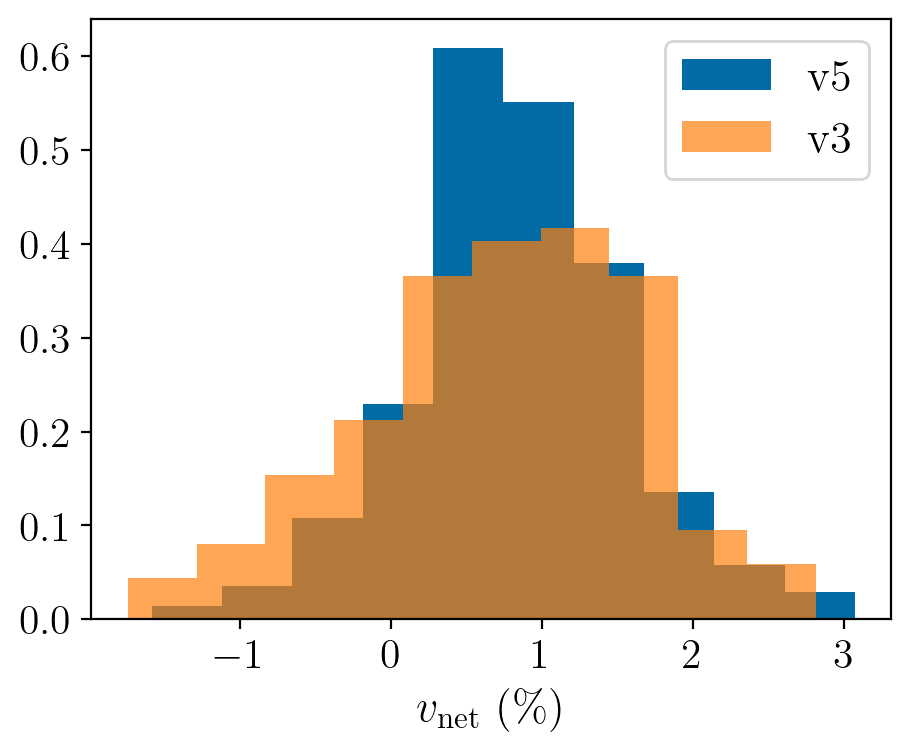}
    \caption{$\vnet$ distributions for a MAD a+0.5, $\rhigh$ 40, inclination $30^\circ$ model with an aligned field configuration for two different GRMHD simulations. "v3" is the simulation set used in this paper, while "v5" is a new image library with a higher GRMHD resolution (384x192x192 compared to 288x128x128 in v3) and different fluid adiabatic index (5/3 instead of 4/3). Both distributions are indistinguishable under a K-S test.}
    \label{fig:v3_v5}
\end{figure}

Changes to GRMHD fluid parameters can influence the resulting $\vnet$ distributions. In a newer set of GRMHD simulations generated using \kharma (referred to as "v5" as opposed to the current "v3"), the simulations are run with a different adiabatic index (5/3 instead of 4/3) for the fluid, different GRMHD floor prescriptions, higher resolution (384x192x192 compared to 288x128x128 in v3) and run out till $50 \times 10^3 GM/c^3$ in time. A comparison between the same GRRT model parameters between this simulation and the simulation used in the paper is given in Fig. \ref{fig:v3_v5} (same model as in section \ref{sec:cp_dist_one}). A 2-sample K-S test cannot distinguish the newer simulation distribution from the one used in this paper and both perform similarly when compared to observational data of Sgr~A*, however this inference cannot be applied to the full library of simulations. Characteristics of the new library, containing densely sampled black hole spins, will be discussed in a later paper.

Electron-positron plasmas have the capacity to greatly influence the morphology of the Stokes V image without affecting the Stokes I image significantly. With an equal proportion of positrons and electrons, both intrinsic emission and Faraday conversion through Faraday rotation vanish and the resultant CP will encode conversion through the sign of twist \citep{wardle_electron-positron_1998,anantua_positron_2020,emami_positron_2021}.

The electron distribution function (eDF) was chosen to be a relativistic Maxwell-Juettner distribution. Based on observations of the solar wind and simulations of collisionless plasma simulations via Particle-in-Cell (PIC) codes, a power-law tail to the distribution can be modeled\textemdash the so-called $\kappa$ distribution \citep[][and references therein]{kunz_2015}. Non-thermal eDFs introduce hotter electrons which influence all the radiative transfer coefficients, however a systematic study of the effects on CP is still needed.

The $\rhigh$ prescription \citep{moscibrodzca16} used to assign electron temperatures in our models is a phenomenological model.  The $\rhigh$ model defines the electron temperatures as a particular function of plasma beta $\beta$.  Another parameterization of the accretion flow is the critical beta model \citep{anantua_critbeta_2020}, which differ in that the electron temperatures approach 0 instead of 1/$\rhigh$ at high plasma beta, in the midplane. Colder electrons in the midplane will enhance Faraday rotation while suppressing intrinsic emission and thus reduce $\vnet$ in the images.

As the observed $\vnet$ for Sgr~A* seem to lie squarely around the -1\% value across many decades, the potential effects of an external Faraday screen should be investigated (though recently argued against by \citet{Wielgus2023}). Faraday conversion and intrinsic emission is heavily suppressed in cold plasmas compared to Faraday rotation, thus the existence of a screen should not greatly affect the $\vnet$ measurements. It is possible, however, for the screen to undergo field reversals along the photon trajectory, in which case Faraday conversion can dominate in a small region where the field is perpendicular to the photon trajectory. \citet{gruzinov_conversion_2019} investigates the effect of such field reversals in cold plasma and find that the resulting CP oscillates quasiperiodically as a function of $\lambda^2$. Since the observations of Sgr~A* do not seem to show this oscillation, we may assume that such effects are subdominant in observations of Sgr~A*. For M87* CP($\lambda$) is not yet observed.

While $\sim 90\%$ of the emission up to $100$mas arises from horizon scales \citepalias{EHTC_2022_2}, emission from large scale structures between $100$mas and $1$as are not as well constrained. The lack of a conclusively observed jet from Sgr~A* at lower frequencies \citepalias[][and references therein]{EHTC_2022_2} also suggest that any extended emission is of low intensity. However, it is possible a highly polarized, low intensity source in this region could influence Stokes V measurements and offset a signal arising from the source. Future improvements to the global VLBI array can improve limits on extended structure \citep{raymond_ngeht_2021}.

\section{Conclusion}

In this paper, we investigate the circular polarization of simulated images of Sgr~A* and M87*, focusing on the image integrated circular polarization ($\vnet$). We explore a library of GRMHD models spanning different black hole spins and accretion disc magnetic states (MAD, SANE). Ray-traced images of the GRMHD models span different electron temperatures, observer inclinations and both aligned and reversed polarities of the global magnetic field. To understand properties of $\vnet$, we first focused on one MAD a+0.5, $\rhigh$ 160, inclination $30^\circ$ model. We then plot $\vnet$ distributions across an entire model library and find trends with respect to the library parameters along with fitting functions for the Sgr~A* MAD models. Models of Sgr~A* are constrained by performing a KS-test between simulations and unresolved ALMA observations of Sgr~A*. We find the following results.

\begin{itemize}
    \item Field reversal does not flip $\vnet$ distributions as there exists both symmetric and antisymmetric terms in the radiative transfer equation (Eq. \ref{eq:StokesV}). The relative contributions of these terms can vary greatly. Models with symmetric (anti-symmetric) terms dominating the equation can have nearly symmetric (anti-symmetric) distributions of $\vnet$. In practice however, most models contain contributions from both terms.
    \item Large cancellations occur both spatially and temporally in CP due to turbulent fluctuations and symmetries in the magnetic field. Average images and means of $\vnet$ distributions can smooth over fluctuations and probe the structure of magnetic fields. The sense of twist of the magnetic field is encoded in the mean of the $\vnet$ distributions (when averaged over field configuration). Inclinations $<90^\circ$ encode an overall clockwise twist of field (positive $\vnet$).
    \item SANE models produce more CP on average than MADs. For Sgr~A*, nearly all the MAD models lie within $|\vnet| < 2\%$ whereas for SANEs $|\vnet| \lesssim 5\%$.
    \item When compared to $\vnet$ ALMA measurements of Sgr~A* via a KS test, MAD models that pass contain $\vb$ pointing away from the observer. All but one of the passing models are clockwise in the sky, in agreement with the direction of the putative orbital motion reported by \citet{gravity_clockwise_2018,wielgus_orbital_2022}. SANE models being more variable with larger $\vnet$ tend to pass without a clear trend. None of the best-bet models survive the $\vnet$ constraint as most of the MAD models exhibit sign changes in $\vnet$ unlike observations which lie closely around the -1\% region.
    \item Edge-on models produce $\vnet \approx 0$ due to symmetries in the magnetic field structure and are thus disfavored for Sgr~A*.
    \item Black hole spin influences the $\vnet$ distribution of a model. High-spin prograde models appear to contain imprints of the photon ring with the opposite sign of CP compared to the weakly lensed component, causing $\vnet$ in prograde models to be centered closer to 0 or even have the opposite sign, compared to otherwise similar retrograde models.
    \item Electron temperature assignment can be constrained by future observations of the resolved CP measurement, $\vavg$. Higher $\rhigh$ models or colder electrons in the disk will have higher $\vavg$ values with $\vnet$ weakly affected.
    
\end{itemize}

Overall, we find that while CP in radiatively inefficient accretion flows can be complicated, there are interesting trends and properties with respect to model parameters. The GRMHD models appear to be highly variable in CP with frequent sign crossings in $\vnet$. Current constraints of $\vnet$ for Sgr~A* seem to highlight the global direction of the magnetic field and the sense of rotation of the flow. Since $\vnet$ observations are possible for point sources, observational data from targets besides Sgr~A* and M87* can also be used to infer model properties.

\acknowledgements
The authors thank Monika Mościbrodzka, Angelo Ricarte and the anonymous EHT internal referee for valuable comments that improved the manuscript.

This work was supported by NSF grants AST 17-16327 (horizon), OISE 17-43747, and AST 20-34306.  This research used resources of the Oak Ridge Leadership 
Computing Facility at the Oak Ridge National Laboratory, which is 
supported by the Office of Science of the U.S. Department of Energy 
under Contract No. DE-AC05-00OR22725.  This research used resources of 
the Argonne Leadership Computing Facility, which is a DOE Office of 
Science User Facility supported under Contract DE-AC02-06CH11357.  This 
research was done using services provided by the OSG Consortium, which is supported by the National Science Foundation awards \#2030508 and \#1836650.
This research is part of the Delta research computing project, which 
is supported by the National Science Foundation (award OCI 2005572), and the State of Illinois. Delta is a joint effort of the University of Illinois at Urbana-Champaign and its National Center for Supercomputing Applications. This paper makes use of the following ALMA data: ADS/JAO.ALMA\#2013.1.00764.S, ADS/JAO.ALMA\#2016.1.01154.V and \newline ADS/JAO.ALMA\#2016.1.01404.V. ALMA is a partnership of ESO (representing its member states), NSF (USA) and NINS (Japan), together with NRC (Canada), NSC and ASIAA (Taiwan), and KASI (Republic of Korea), in cooperation with the Republic of Chile. The National Radio Astronomy Observatory is a facility of the National Science Foundation operated under cooperative agreement by Associated Universities, Inc. CFG was supported by the IBM Einstein Fellow Fund at the Institute for Advanced Study. MW acknowledges the support by the European Research Council advanced grant “M2FINDERS - Mapping Magnetic Fields with INterferometry Down to Event hoRizon Scales” (Grant No. 101018682).

\appendix

\renewcommand{\thesection}{\Alph{section}}

\section{One-zone Model Stokes $V$ vs Frequency}\label{app:one_zone_cp_estimate}
In the optically/Faraday thin limit, we can approximate the contributing terms (intrinsic emission and Faraday conversion) to Stokes $V$ from the full solution of the radiative transfer equation (eq. \ref{eq:GenRadTrans}) as follows.
\begin{equation}\label{eq:v_emission_proxy}
    V_\mathrm{emission} \approx j_V L
\end{equation}
\begin{equation}\label{eq:v_conversion_proxy}
    V_\mathrm{conversion} \approx \rho_Q \rho_V j_Q L^3
\end{equation}
Where Stokes $V$ from Faraday conversion in a uniform field geometry only arises from Faraday conversion of linearly polarized light rotated from $Q$ to $U$. $L$ is a characteristic length-scale which we set to the radius of the one-zone sphere $L=5 G M / c^2$. Thus Eq. \ref{eq:v_emission_proxy} and \ref{eq:v_conversion_proxy} can be used as proxies to probe the general spectral behavior of the components to the full solution of Stokes V.

\begin{figure}
    \centering
    \includegraphics[width=\linewidth]{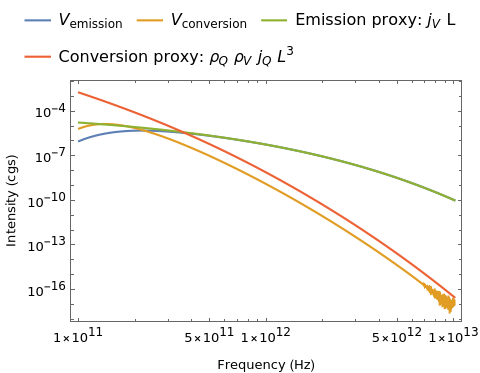}
    \caption{Stokes $V$ vs frequency for solutions to the radiative transfer equation consisting of either only intrinsic emission ($V_\mathrm{emission}$) or Faraday conversion ($V_\mathrm{conversion}$), in the Faraday thin limit. Estimates to these solutions (proxies) from dimensional analysis of the components are also plotted and display the same scaling behavior as the corresponding solution.}
    \label{fig:stokesv_analytic_freq}
\end{figure}

Figure \ref{fig:stokesv_analytic_freq} shows the spectral behavior of the two CP mechanisms along with their approximations in the Faraday thin limit using one-zone models of Sgr~A*. The Stokes $V_\mathrm{emission}$ ($V_\mathrm{conversion}$) solution is obtained by setting $\rho_Q=0$ ($j_V=\alpha_V=0$) respectively. The approximations have the same scaling behavior as the full solutions, with CP from Faraday conversion decreasing much faster with frequency than intrinsic emission.

When comparing with numerical models, we see the same qualitative behavior: models in which intrinsic emission is expected to dominate show a slower decrease in frequency compared to conversion dominated models. The spectral slopes of Stokes $V$ are completely different, however this is not unexpected given the nontrivial field configuration of the simulations and the integration across many different geodesics (compared to a single geodesic in the analytic solution).

\section{Validity of Fractional CP Distribution}\label{app:cp_convergence}

Before making estimates and predictions from the surveyed data, it should be tested that the distributions computed from the EHT imaging library have converged to the true distribution of the process. Here we investigate convergence properties of the models with respect to resolution in time, GRMHD and GRRT modeling.

\begin{figure}
    \centering
    \includegraphics[width=0.95\linewidth]{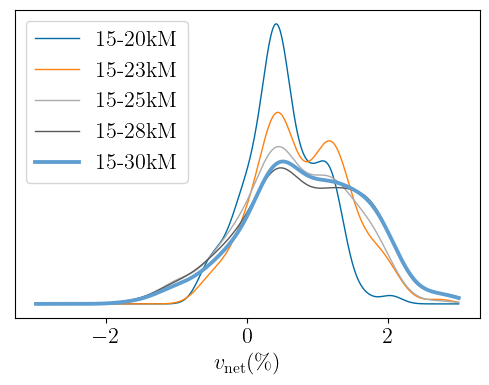}
    \caption{CP distributions for a Sgr~A* MAD, spin +0.5, $\rhigh$ 160, inclination $30^\circ$ model for larger time ranges. While the distribution has not fully converged, the fluctuations in the mean are relatively low. Mean values for each distribution with increasing time range are: 0.51\%, 0.78\%, 0.67\%, 0.79\% and 0.82\% respectively. The total distribution used for analyses is given by the thick, light-blue line.}
    \label{fig:CP_timescale_comparison}
\end{figure}

Have the distributions converged i.e., has the library been imaged over a long enough interval? Fig. \ref{fig:CP_timescale_comparison} displays the effect of increasing the simulation time for our sample MAD, spin +0.5, $\rhigh$ 160, inclination $30^\circ$ model. While the mean of the distribution is not fully settled, changes are within 10\%: from 5kM to 15kM time intervals, in increments of 3kM, the mean $\vnet$ changes as 0.51\%, 0.78\%, 0.67\%, 0.79\% and 0.82\% respectively. The distributions also become more unimodal as the number of independent samples increases. The correlation time of this model is about 400M, implying about 38 independent samples and a standard error of $\sigma / \sqrt{n} \approx 0.13\%$.

\begin{figure}
    \centering
    \includegraphics[width=0.95\linewidth]{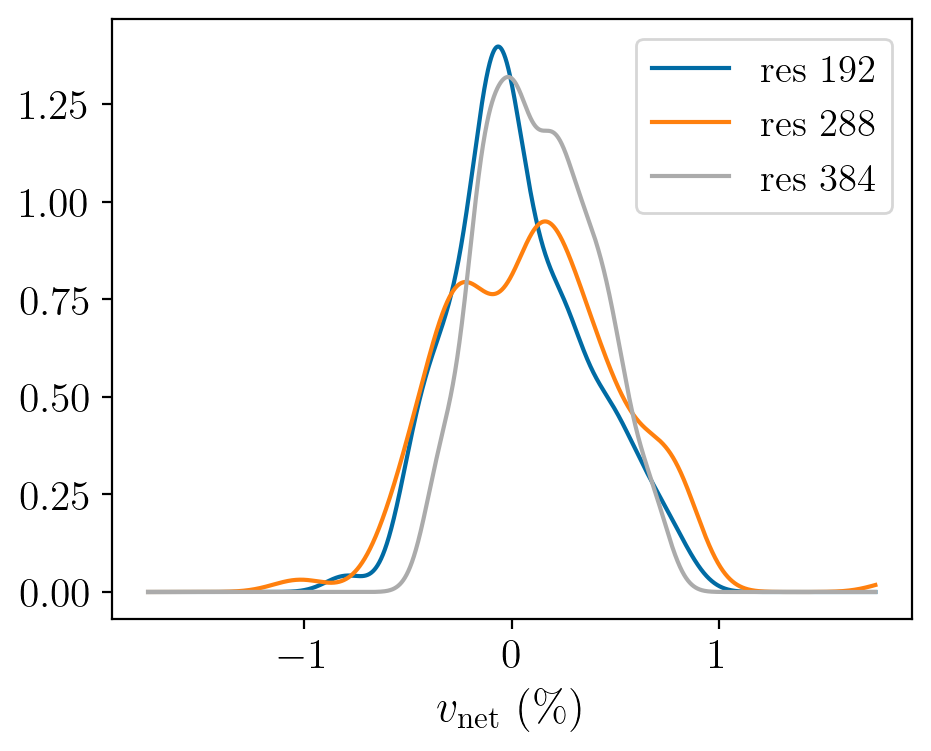}
    \caption{Kernel density estimation PDFs for different GRMHD simulation resolutions for a spin +0.94 MAD model with M87 parameters. The full resolution of each of the distribution (radial x poloidal x azimuthal) is 192x96x96, 288x128x128 and 384x192x192 respectively. This demonstrates that resolution does not influence the distributions significantly.}
    \label{fig:resolution_check}
\end{figure}

The distributions might have encoded features dependent on resolution of the GRMHD simulations. Fig. \ref{fig:resolution_check} shows that this is not the case for a MAD spin +0.94 model, as the resolution does not drastically affect the distribution. The GRMHD resolution used for the EHT imaging in \citetalias{EHTC_2022_5} and this paper was 288x128x128. The fluctuations in the distributions are likely due to limited time sampling of the model (both in cadence and length).

$\vnet$ is a metric that is independent of the image (GRRT) resolution past a certain threshold which represents the scale of critical structures in the image. We generate a snapshot for a MAD spin $+0.94$ model at resolutions 0.25, 0.5, 1, 2, 4, 8 and 16 $\mu$as/pixel and find the $\vnet$ values to be 0.361\%, 0.358\%, 0.340\%, 0.384\%, 0.476\%, 0.447\% and 0.335\% respectively. Downsampling the model library (0.5$\mu$as/pixel) across all parameters shows similar trends suggesting that $\vnet$ measurements of GRRT models usually vary within 0.1\% across image resolution until 8 $\mu$as/pixel, making the existing image resolution of the library sufficient.

\section{M87 $\vnet$ Plots}\label{app:m87_vnet}

\begin{figure*}
    \centering
    \includegraphics[width=\textwidth]{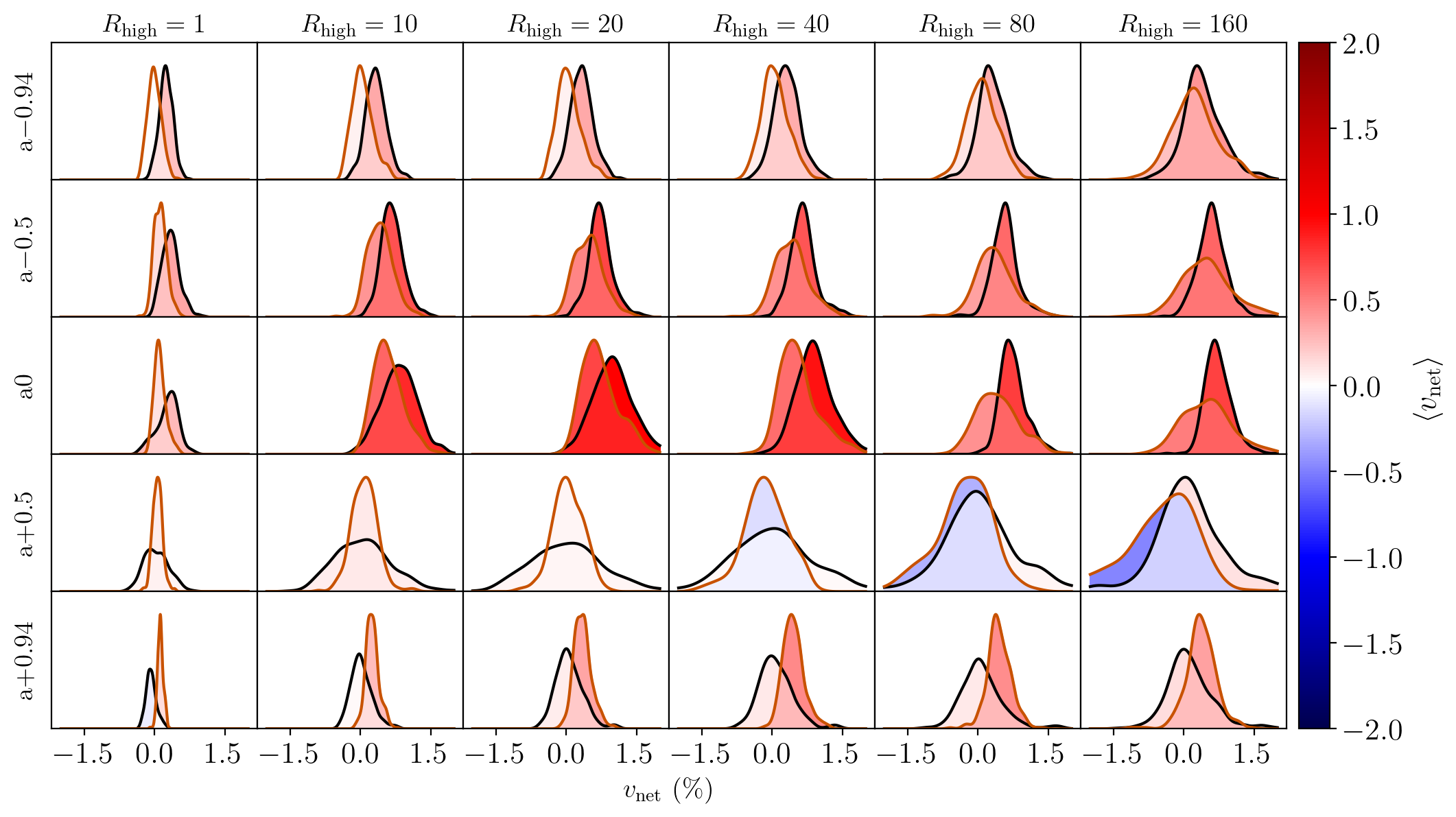}
    \caption{Distributions of $\vnet$ for all the MAD M87 models. Y-axis corresponds to spin. X-axis corresponds to $\rhigh$. Inclination is $163^\circ$ for $a_* \geq 0$ and $17^\circ$ otherwise. The black (orange) lines represent the aligned (reversed) field distribution respectively. The color filled within the distributions is their mean $\vnet$ and the overlapping regions is the mean $\vnet$ of both field distributions combined. The height of each subplot is adjusted to fix the maximum height constant for visualization purposes.}
    \label{fig:M87_distributions_MAD}
\end{figure*}

\begin{figure*}
    \centering
    \includegraphics[width=\textwidth]{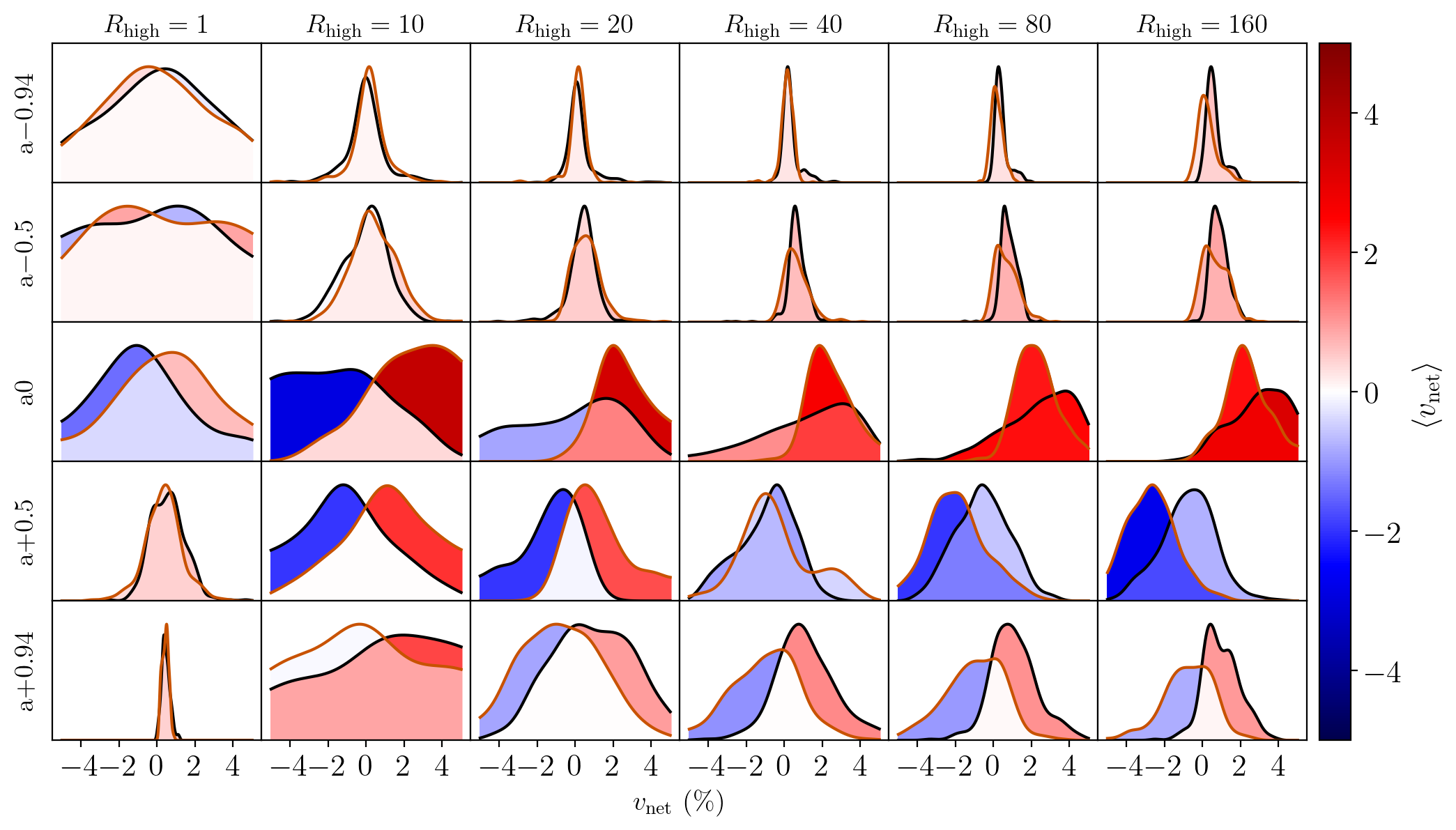}
    \caption{Same as Fig. \ref{fig:M87_distributions_MAD} except for SANE distributions.}
    \label{fig:M87_distributions_SANE}
\end{figure*}

Figures \ref{fig:M87_distributions_MAD}, \ref{fig:M87_distributions_SANE} show $\vnet$ distributions for the M87* library parameters. As the observer inclination for the M87* library is fixed to the inclination angle of the forward jet, these M87* distributions are mostly a subset of the parameters used for Sgr~A* aside from a denser sampling of $\rhigh$. Thus inferences on the parameter trends based on the Sgr~A* distributions also apply to M87*: SANEs produce more $\vnet$ than MAD models, the shift of $\vnet$ distributions goes towards 0 as spin goes from negative to positive and an increase in $\rhigh$ increases and broadens the $\vnet$ distributions.

\section{CP Distribution Tables}\label{appsec:cp_tables}

Here we provide tables of the first two moments of the distributions for $\vnet$(\%) for both Sgr~A* and M87* in Table \ref{tab:moments}.

\setlength\tabcolsep{6pt}
\begin{longtable*}[p]{c | c | c | c | c | c | c | c | c }
\caption{Mean and standard deviation for $\vnet$(\%) (fractional CP \%) for all GRMHD models for SgrA\textsuperscript{*} and M87\textsuperscript{*}.}\label{tab:moments}\\

Source & Flux & a & $R_\mathrm{high}$ & $\theta$ & Mean $\vnet$ & Std $\vnet$ & Mean $\vnet$ (Rev $\vb$) & Std. $\vnet$ (Rev $\vb$)\\
\hline
\endfirsthead

Source & Flux & a & $R_\mathrm{high}$ & $\theta$ & Mean $\vnet$ & Std $\vnet$ & Mean $\vnet$ (Rev $\vb$) & Std. $\vnet$ (Rev $\vb$)\\
\hline
\endhead

\hline
\endfoot

\hline \hline
\endlastfoot
    
SgrA\textsuperscript{*} & SANE & $-0.94$ & 1 & 10 & 0.62 & 2.80 & -0.15 & 2.85 \\ 
SgrA\textsuperscript{*} & SANE & $-0.94$ & 1 & 30 & -0.71 & 1.32 & -1.00 & 1.36 \\ 
SgrA\textsuperscript{*} & SANE & $-0.94$ & 1 & 50 & -1.61 & 1.96 & -1.96 & 2.01 \\ 
SgrA\textsuperscript{*} & SANE & $-0.94$ & 1 & 70 & -1.66 & 2.38 & -2.16 & 2.49 \\ 
SgrA\textsuperscript{*} & SANE & $-0.94$ & 1 & 90 & 0.32 & 2.04 & -0.36 & 2.11 \\ 
SgrA\textsuperscript{*} & SANE & $-0.94$ & 1 & 110 & 2.19 & 2.37 & 1.56 & 2.26 \\ 
SgrA\textsuperscript{*} & SANE & $-0.94$ & 1 & 130 & 1.72 & 2.01 & 1.55 & 1.91 \\ 
SgrA\textsuperscript{*} & SANE & $-0.94$ & 1 & 150 & 0.55 & 1.31 & 0.69 & 1.28 \\ 
SgrA\textsuperscript{*} & SANE & $-0.94$ & 1 & 170 & -0.34 & 2.73 & -0.65 & 2.62 \\ 
SgrA\textsuperscript{*} & SANE & $-0.94$ & 10 & 10 & -0.62 & 2.87 & 0.65 & 2.79 \\ 
SgrA\textsuperscript{*} & SANE & $-0.94$ & 10 & 30 & -0.55 & 1.81 & 0.75 & 1.85 \\ 
SgrA\textsuperscript{*} & SANE & $-0.94$ & 10 & 50 & -0.28 & 1.00 & 0.35 & 0.98 \\ 
SgrA\textsuperscript{*} & SANE & $-0.94$ & 10 & 70 & -0.16 & 0.73 & 0.09 & 0.75 \\ 
SgrA\textsuperscript{*} & SANE & $-0.94$ & 10 & 90 & -0.22 & 0.64 & 0.19 & 0.64 \\ 
SgrA\textsuperscript{*} & SANE & $-0.94$ & 10 & 110 & -0.23 & 0.73 & 0.25 & 0.70 \\ 
SgrA\textsuperscript{*} & SANE & $-0.94$ & 10 & 130 & -0.31 & 1.00 & 0.26 & 1.02 \\ 
SgrA\textsuperscript{*} & SANE & $-0.94$ & 10 & 150 & -0.88 & 1.84 & 0.69 & 1.81 \\ 
SgrA\textsuperscript{*} & SANE & $-0.94$ & 10 & 170 & -1.02 & 2.22 & 0.97 & 2.21 \\ 
SgrA\textsuperscript{*} & SANE & $-0.94$ & 40 & 10 & 0.55 & 1.49 & -0.32 & 1.35 \\ 
SgrA\textsuperscript{*} & SANE & $-0.94$ & 40 & 30 & 0.46 & 1.03 & 0.26 & 0.94 \\ 
SgrA\textsuperscript{*} & SANE & $-0.94$ & 40 & 50 & 0.28 & 0.76 & 0.22 & 0.70 \\ 
SgrA\textsuperscript{*} & SANE & $-0.94$ & 40 & 70 & 0.20 & 0.63 & -0.03 & 0.62 \\ 
SgrA\textsuperscript{*} & SANE & $-0.94$ & 40 & 90 & -0.10 & 0.53 & 0.14 & 0.54 \\ 
SgrA\textsuperscript{*} & SANE & $-0.94$ & 40 & 110 & -0.37 & 0.59 & 0.28 & 0.60 \\ 
SgrA\textsuperscript{*} & SANE & $-0.94$ & 40 & 130 & -0.58 & 0.78 & 0.27 & 0.72 \\ 
SgrA\textsuperscript{*} & SANE & $-0.94$ & 40 & 150 & -1.01 & 1.16 & 0.56 & 1.08 \\ 
SgrA\textsuperscript{*} & SANE & $-0.94$ & 40 & 170 & -1.32 & 1.11 & 1.17 & 1.14 \\ 
SgrA\textsuperscript{*} & SANE & $-0.94$ & 160 & 10 & 0.89 & 0.54 & -0.36 & 0.53 \\ 
SgrA\textsuperscript{*} & SANE & $-0.94$ & 160 & 30 & 1.00 & 0.48 & -0.14 & 0.48 \\ 
SgrA\textsuperscript{*} & SANE & $-0.94$ & 160 & 50 & 0.78 & 0.50 & -0.09 & 0.54 \\ 
SgrA\textsuperscript{*} & SANE & $-0.94$ & 160 & 70 & 0.39 & 0.45 & -0.16 & 0.49 \\ 
SgrA\textsuperscript{*} & SANE & $-0.94$ & 160 & 90 & 0.04 & 0.41 & -0.16 & 0.46 \\ 
SgrA\textsuperscript{*} & SANE & $-0.94$ & 160 & 110 & -0.24 & 0.42 & -0.20 & 0.51 \\ 
SgrA\textsuperscript{*} & SANE & $-0.94$ & 160 & 130 & -0.77 & 0.37 & -0.11 & 0.46 \\ 
SgrA\textsuperscript{*} & SANE & $-0.94$ & 160 & 150 & -0.91 & 0.40 & 0.14 & 0.43 \\ 
SgrA\textsuperscript{*} & SANE & $-0.94$ & 160 & 170 & -0.83 & 0.38 & 0.46 & 0.41 \\ 
SgrA\textsuperscript{*} & SANE & $-0.5$ & 1 & 10 & 0.07 & 1.02 & 0.49 & 1.02 \\ 
SgrA\textsuperscript{*} & SANE & $-0.5$ & 1 & 30 & -0.59 & 0.57 & -0.58 & 0.52 \\ 
SgrA\textsuperscript{*} & SANE & $-0.5$ & 1 & 50 & -1.86 & 1.12 & -2.23 & 1.18 \\ 
SgrA\textsuperscript{*} & SANE & $-0.5$ & 1 & 70 & -2.55 & 1.62 & -3.25 & 1.71 \\ 
SgrA\textsuperscript{*} & SANE & $-0.5$ & 1 & 90 & 0.25 & 2.09 & -0.54 & 1.91 \\ 
SgrA\textsuperscript{*} & SANE & $-0.5$ & 1 & 110 & 3.18 & 1.67 & 2.81 & 1.77 \\ 
SgrA\textsuperscript{*} & SANE & $-0.5$ & 1 & 130 & 2.32 & 1.22 & 1.99 & 1.16 \\ 
SgrA\textsuperscript{*} & SANE & $-0.5$ & 1 & 150 & 0.70 & 0.60 & 0.69 & 0.54 \\ 
SgrA\textsuperscript{*} & SANE & $-0.5$ & 1 & 170 & -0.38 & 0.89 & 0.03 & 0.90 \\ 
SgrA\textsuperscript{*} & SANE & $-0.5$ & 10 & 10 & -0.43 & 2.65 & 0.45 & 2.62 \\ 
SgrA\textsuperscript{*} & SANE & $-0.5$ & 10 & 30 & 0.05 & 2.17 & 0.17 & 2.16 \\ 
SgrA\textsuperscript{*} & SANE & $-0.5$ & 10 & 50 & 0.04 & 1.11 & -0.02 & 1.08 \\ 
SgrA\textsuperscript{*} & SANE & $-0.5$ & 10 & 70 & -0.03 & 0.70 & -0.06 & 0.72 \\ 
SgrA\textsuperscript{*} & SANE & $-0.5$ & 10 & 90 & -0.02 & 0.65 & -0.06 & 0.63 \\ 
SgrA\textsuperscript{*} & SANE & $-0.5$ & 10 & 110 & 0.01 & 0.79 & -0.06 & 0.73 \\ 
SgrA\textsuperscript{*} & SANE & $-0.5$ & 10 & 130 & -0.02 & 0.99 & -0.15 & 1.00 \\ 
SgrA\textsuperscript{*} & SANE & $-0.5$ & 10 & 150 & -0.23 & 2.07 & -0.17 & 1.99 \\ 
SgrA\textsuperscript{*} & SANE & $-0.5$ & 10 & 170 & -0.21 & 3.46 & 0.13 & 3.36 \\ 
SgrA\textsuperscript{*} & SANE & $-0.5$ & 40 & 10 & 0.38 & 0.81 & 0.13 & 0.86 \\ 
SgrA\textsuperscript{*} & SANE & $-0.5$ & 40 & 30 & 0.71 & 0.59 & -0.04 & 0.63 \\ 
SgrA\textsuperscript{*} & SANE & $-0.5$ & 40 & 50 & 0.51 & 0.55 & -0.06 & 0.58 \\ 
SgrA\textsuperscript{*} & SANE & $-0.5$ & 40 & 70 & 0.21 & 0.48 & -0.14 & 0.57 \\ 
SgrA\textsuperscript{*} & SANE & $-0.5$ & 40 & 90 & -0.14 & 0.46 & -0.17 & 0.55 \\ 
SgrA\textsuperscript{*} & SANE & $-0.5$ & 40 & 110 & -0.55 & 0.55 & 0.02 & 0.53 \\ 
SgrA\textsuperscript{*} & SANE & $-0.5$ & 40 & 130 & -1.00 & 0.65 & 0.10 & 0.61 \\ 
SgrA\textsuperscript{*} & SANE & $-0.5$ & 40 & 150 & -1.20 & 0.75 & -0.01 & 0.78 \\ 
SgrA\textsuperscript{*} & SANE & $-0.5$ & 40 & 170 & -0.83 & 0.98 & -0.06 & 1.00 \\ 
SgrA\textsuperscript{*} & SANE & $-0.5$ & 160 & 10 & 1.31 & 0.45 & 0.07 & 0.72 \\ 
SgrA\textsuperscript{*} & SANE & $-0.5$ & 160 & 30 & 1.66 & 0.44 & 0.23 & 0.68 \\ 
SgrA\textsuperscript{*} & SANE & $-0.5$ & 160 & 50 & 1.24 & 0.49 & 0.09 & 0.61 \\ 
SgrA\textsuperscript{*} & SANE & $-0.5$ & 160 & 70 & 0.70 & 0.48 & -0.04 & 0.61 \\ 
SgrA\textsuperscript{*} & SANE & $-0.5$ & 160 & 90 & 0.07 & 0.39 & -0.13 & 0.61 \\ 
SgrA\textsuperscript{*} & SANE & $-0.5$ & 160 & 110 & -0.60 & 0.42 & -0.10 & 0.59 \\ 
SgrA\textsuperscript{*} & SANE & $-0.5$ & 160 & 130 & -1.32 & 0.44 & -0.09 & 0.62 \\ 
SgrA\textsuperscript{*} & SANE & $-0.5$ & 160 & 150 & -1.70 & 0.47 & -0.28 & 0.65 \\ 
SgrA\textsuperscript{*} & SANE & $-0.5$ & 160 & 170 & -1.33 & 0.56 & -0.18 & 0.60 \\ 
SgrA\textsuperscript{*} & SANE & $0$ & 1 & 10 & -0.06 & 0.29 & 0.11 & 0.26 \\ 
SgrA\textsuperscript{*} & SANE & $0$ & 1 & 30 & -0.49 & 0.25 & -0.42 & 0.21 \\ 
SgrA\textsuperscript{*} & SANE & $0$ & 1 & 50 & -1.42 & 0.42 & -1.51 & 0.47 \\ 
SgrA\textsuperscript{*} & SANE & $0$ & 1 & 70 & -1.78 & 0.63 & -2.25 & 0.64 \\ 
SgrA\textsuperscript{*} & SANE & $0$ & 1 & 90 & 0.43 & 1.02 & -0.31 & 0.93 \\ 
SgrA\textsuperscript{*} & SANE & $0$ & 1 & 110 & 2.13 & 0.55 & 2.08 & 0.52 \\ 
SgrA\textsuperscript{*} & SANE & $0$ & 1 & 130 & 1.46 & 0.41 & 1.46 & 0.43 \\ 
SgrA\textsuperscript{*} & SANE & $0$ & 1 & 150 & 0.35 & 0.23 & 0.48 & 0.20 \\ 
SgrA\textsuperscript{*} & SANE & $0$ & 1 & 170 & -0.18 & 0.31 & 0.02 & 0.26 \\ 
SgrA\textsuperscript{*} & SANE & $0$ & 10 & 10 & -2.30 & 3.54 & 2.51 & 3.59 \\ 
SgrA\textsuperscript{*} & SANE & $0$ & 10 & 30 & -1.05 & 2.17 & 1.26 & 2.29 \\ 
SgrA\textsuperscript{*} & SANE & $0$ & 10 & 50 & -0.23 & 1.08 & 0.06 & 1.13 \\ 
SgrA\textsuperscript{*} & SANE & $0$ & 10 & 70 & -0.32 & 0.73 & 0.06 & 0.78 \\ 
SgrA\textsuperscript{*} & SANE & $0$ & 10 & 90 & -0.25 & 0.70 & 0.28 & 0.69 \\ 
SgrA\textsuperscript{*} & SANE & $0$ & 10 & 110 & 0.05 & 0.80 & 0.37 & 0.66 \\ 
SgrA\textsuperscript{*} & SANE & $0$ & 10 & 130 & -0.37 & 1.07 & 0.70 & 1.11 \\ 
SgrA\textsuperscript{*} & SANE & $0$ & 10 & 150 & -1.71 & 2.31 & 1.57 & 2.35 \\ 
SgrA\textsuperscript{*} & SANE & $0$ & 10 & 170 & -1.51 & 3.50 & 1.44 & 3.47 \\ 
SgrA\textsuperscript{*} & SANE & $0$ & 40 & 10 & -0.99 & 2.84 & 3.36 & 1.70 \\ 
SgrA\textsuperscript{*} & SANE & $0$ & 40 & 30 & 0.14 & 1.59 & 1.33 & 1.09 \\ 
SgrA\textsuperscript{*} & SANE & $0$ & 40 & 50 & 0.17 & 1.00 & 0.27 & 0.78 \\ 
SgrA\textsuperscript{*} & SANE & $0$ & 40 & 70 & -0.31 & 0.95 & 0.08 & 0.65 \\ 
SgrA\textsuperscript{*} & SANE & $0$ & 40 & 90 & -0.23 & 0.88 & 0.27 & 0.60 \\ 
SgrA\textsuperscript{*} & SANE & $0$ & 40 & 110 & -0.27 & 0.88 & 0.88 & 0.86 \\ 
SgrA\textsuperscript{*} & SANE & $0$ & 40 & 130 & -1.24 & 0.90 & 1.23 & 0.92 \\ 
SgrA\textsuperscript{*} & SANE & $0$ & 40 & 150 & -2.08 & 1.10 & 1.03 & 1.17 \\ 
SgrA\textsuperscript{*} & SANE & $0$ & 40 & 170 & -2.06 & 1.30 & 0.12 & 1.51 \\ 
SgrA\textsuperscript{*} & SANE & $0$ & 160 & 10 & 3.00 & 1.79 & 1.88 & 1.22 \\ 
SgrA\textsuperscript{*} & SANE & $0$ & 160 & 30 & 2.45 & 1.10 & 1.00 & 1.09 \\ 
SgrA\textsuperscript{*} & SANE & $0$ & 160 & 50 & 1.74 & 0.83 & 0.02 & 0.95 \\ 
SgrA\textsuperscript{*} & SANE & $0$ & 160 & 70 & 0.72 & 1.02 & -0.18 & 0.95 \\ 
SgrA\textsuperscript{*} & SANE & $0$ & 160 & 90 & -0.08 & 1.05 & -0.10 & 0.84 \\ 
SgrA\textsuperscript{*} & SANE & $0$ & 160 & 110 & -0.92 & 0.81 & 0.19 & 0.96 \\ 
SgrA\textsuperscript{*} & SANE & $0$ & 160 & 130 & -1.77 & 0.55 & -0.02 & 0.89 \\ 
SgrA\textsuperscript{*} & SANE & $0$ & 160 & 150 & -2.61 & 0.59 & -0.68 & 0.99 \\ 
SgrA\textsuperscript{*} & SANE & $0$ & 160 & 170 & -3.46 & 0.94 & -1.37 & 1.12 \\ 
SgrA\textsuperscript{*} & SANE & $+0.5$ & 1 & 10 & -0.01 & 0.17 & -0.01 & 0.18 \\ 
SgrA\textsuperscript{*} & SANE & $+0.5$ & 1 & 30 & -0.37 & 0.13 & -0.36 & 0.14 \\ 
SgrA\textsuperscript{*} & SANE & $+0.5$ & 1 & 50 & -1.09 & 0.24 & -1.11 & 0.24 \\ 
SgrA\textsuperscript{*} & SANE & $+0.5$ & 1 & 70 & -1.26 & 0.37 & -1.41 & 0.31 \\ 
SgrA\textsuperscript{*} & SANE & $+0.5$ & 1 & 90 & -0.04 & 0.42 & -0.04 & 0.44 \\ 
SgrA\textsuperscript{*} & SANE & $+0.5$ & 1 & 110 & 1.41 & 0.33 & 1.46 & 0.36 \\ 
SgrA\textsuperscript{*} & SANE & $+0.5$ & 1 & 130 & 1.16 & 0.30 & 1.20 & 0.27 \\ 
SgrA\textsuperscript{*} & SANE & $+0.5$ & 1 & 150 & 0.38 & 0.16 & 0.39 & 0.14 \\ 
SgrA\textsuperscript{*} & SANE & $+0.5$ & 1 & 170 & -0.00 & 0.17 & -0.02 & 0.18 \\ 
SgrA\textsuperscript{*} & SANE & $+0.5$ & 10 & 10 & -1.00 & 4.18 & 1.05 & 4.26 \\ 
SgrA\textsuperscript{*} & SANE & $+0.5$ & 10 & 30 & -1.17 & 2.22 & -0.01 & 2.16 \\ 
SgrA\textsuperscript{*} & SANE & $+0.5$ & 10 & 50 & -0.40 & 1.12 & -0.84 & 1.20 \\ 
SgrA\textsuperscript{*} & SANE & $+0.5$ & 10 & 70 & -0.33 & 0.66 & -0.38 & 0.72 \\ 
SgrA\textsuperscript{*} & SANE & $+0.5$ & 10 & 90 & -0.28 & 0.69 & 0.09 & 0.63 \\ 
SgrA\textsuperscript{*} & SANE & $+0.5$ & 10 & 110 & 0.28 & 0.70 & 0.71 & 0.71 \\ 
SgrA\textsuperscript{*} & SANE & $+0.5$ & 10 & 130 & 0.34 & 1.08 & 1.19 & 1.17 \\ 
SgrA\textsuperscript{*} & SANE & $+0.5$ & 10 & 150 & -0.41 & 2.30 & 1.50 & 2.38 \\ 
SgrA\textsuperscript{*} & SANE & $+0.5$ & 10 & 170 & -0.40 & 4.13 & 0.54 & 4.14 \\ 
SgrA\textsuperscript{*} & SANE & $+0.5$ & 40 & 10 & -2.60 & 3.42 & 2.80 & 3.39 \\ 
SgrA\textsuperscript{*} & SANE & $+0.5$ & 40 & 30 & -0.95 & 1.45 & 0.99 & 1.47 \\ 
SgrA\textsuperscript{*} & SANE & $+0.5$ & 40 & 50 & -0.63 & 0.93 & 0.18 & 1.01 \\ 
SgrA\textsuperscript{*} & SANE & $+0.5$ & 40 & 70 & -0.83 & 0.59 & -0.25 & 0.61 \\ 
SgrA\textsuperscript{*} & SANE & $+0.5$ & 40 & 90 & -0.51 & 0.57 & 0.28 & 0.49 \\ 
SgrA\textsuperscript{*} & SANE & $+0.5$ & 40 & 110 & -0.04 & 0.68 & 0.97 & 0.56 \\ 
SgrA\textsuperscript{*} & SANE & $+0.5$ & 40 & 130 & -0.65 & 0.78 & 1.35 & 0.82 \\ 
SgrA\textsuperscript{*} & SANE & $+0.5$ & 40 & 150 & -1.49 & 1.41 & 1.66 & 1.50 \\ 
SgrA\textsuperscript{*} & SANE & $+0.5$ & 40 & 170 & -1.48 & 1.59 & 1.53 & 1.59 \\ 
SgrA\textsuperscript{*} & SANE & $+0.5$ & 160 & 10 & 0.25 & 1.11 & 1.48 & 1.36 \\ 
SgrA\textsuperscript{*} & SANE & $+0.5$ & 160 & 30 & 0.98 & 0.66 & -0.34 & 0.93 \\ 
SgrA\textsuperscript{*} & SANE & $+0.5$ & 160 & 50 & 0.28 & 0.53 & -1.18 & 0.86 \\ 
SgrA\textsuperscript{*} & SANE & $+0.5$ & 160 & 70 & -0.63 & 0.74 & -1.21 & 0.70 \\ 
SgrA\textsuperscript{*} & SANE & $+0.5$ & 160 & 90 & -0.28 & 0.66 & -0.00 & 0.63 \\ 
SgrA\textsuperscript{*} & SANE & $+0.5$ & 160 & 110 & 0.44 & 0.68 & 1.09 & 0.67 \\ 
SgrA\textsuperscript{*} & SANE & $+0.5$ & 160 & 130 & -0.13 & 0.57 & 1.18 & 0.80 \\ 
SgrA\textsuperscript{*} & SANE & $+0.5$ & 160 & 150 & -0.46 & 0.87 & 0.50 & 0.98 \\ 
SgrA\textsuperscript{*} & SANE & $+0.5$ & 160 & 170 & 0.53 & 1.37 & -1.20 & 1.36 \\ 
SgrA\textsuperscript{*} & SANE & $+0.94$ & 1 & 10 & -0.05 & 0.11 & -0.03 & 0.10 \\ 
SgrA\textsuperscript{*} & SANE & $+0.94$ & 1 & 30 & -0.20 & 0.09 & -0.21 & 0.10 \\ 
SgrA\textsuperscript{*} & SANE & $+0.94$ & 1 & 50 & -0.46 & 0.16 & -0.53 & 0.15 \\ 
SgrA\textsuperscript{*} & SANE & $+0.94$ & 1 & 70 & -0.46 & 0.21 & -0.57 & 0.17 \\ 
SgrA\textsuperscript{*} & SANE & $+0.94$ & 1 & 90 & 0.00 & 0.16 & 0.00 & 0.16 \\ 
SgrA\textsuperscript{*} & SANE & $+0.94$ & 1 & 110 & 0.45 & 0.18 & 0.54 & 0.17 \\ 
SgrA\textsuperscript{*} & SANE & $+0.94$ & 1 & 130 & 0.47 & 0.13 & 0.52 & 0.14 \\ 
SgrA\textsuperscript{*} & SANE & $+0.94$ & 1 & 150 & 0.19 & 0.09 & 0.22 & 0.07 \\ 
SgrA\textsuperscript{*} & SANE & $+0.94$ & 1 & 170 & 0.03 & 0.12 & 0.05 & 0.10 \\ 
SgrA\textsuperscript{*} & SANE & $+0.94$ & 10 & 10 & -1.03 & 2.66 & -1.55 & 2.42 \\ 
SgrA\textsuperscript{*} & SANE & $+0.94$ & 10 & 30 & -2.86 & 1.45 & -3.49 & 1.33 \\ 
SgrA\textsuperscript{*} & SANE & $+0.94$ & 10 & 50 & -4.24 & 1.10 & -4.61 & 1.43 \\ 
SgrA\textsuperscript{*} & SANE & $+0.94$ & 10 & 70 & -2.16 & 1.14 & -3.00 & 1.30 \\ 
SgrA\textsuperscript{*} & SANE & $+0.94$ & 10 & 90 & 0.07 & 0.88 & -0.20 & 0.77 \\ 
SgrA\textsuperscript{*} & SANE & $+0.94$ & 10 & 110 & 2.16 & 1.18 & 2.93 & 1.43 \\ 
SgrA\textsuperscript{*} & SANE & $+0.94$ & 10 & 130 & 4.45 & 1.35 & 4.79 & 1.30 \\ 
SgrA\textsuperscript{*} & SANE & $+0.94$ & 10 & 150 & 3.20 & 1.48 & 3.32 & 1.44 \\ 
SgrA\textsuperscript{*} & SANE & $+0.94$ & 10 & 170 & 1.42 & 2.56 & 1.18 & 2.59 \\ 
SgrA\textsuperscript{*} & SANE & $+0.94$ & 40 & 10 & -0.09 & 0.88 & -0.03 & 0.89 \\ 
SgrA\textsuperscript{*} & SANE & $+0.94$ & 40 & 30 & 0.30 & 0.56 & -0.65 & 0.61 \\ 
SgrA\textsuperscript{*} & SANE & $+0.94$ & 40 & 50 & -0.09 & 0.44 & -0.85 & 0.47 \\ 
SgrA\textsuperscript{*} & SANE & $+0.94$ & 40 & 70 & -0.48 & 0.47 & -0.76 & 0.45 \\ 
SgrA\textsuperscript{*} & SANE & $+0.94$ & 40 & 90 & 0.05 & 0.56 & -0.02 & 0.41 \\ 
SgrA\textsuperscript{*} & SANE & $+0.94$ & 40 & 110 & 0.59 & 0.47 & 0.71 & 0.44 \\ 
SgrA\textsuperscript{*} & SANE & $+0.94$ & 40 & 130 & 0.15 & 0.43 & 0.81 & 0.50 \\ 
SgrA\textsuperscript{*} & SANE & $+0.94$ & 40 & 150 & -0.27 & 0.60 & 0.59 & 0.72 \\ 
SgrA\textsuperscript{*} & SANE & $+0.94$ & 40 & 170 & 0.06 & 1.00 & 0.02 & 1.04 \\ 
SgrA\textsuperscript{*} & SANE & $+0.94$ & 160 & 10 & -0.66 & 0.98 & 0.45 & 1.11 \\ 
SgrA\textsuperscript{*} & SANE & $+0.94$ & 160 & 30 & 0.29 & 0.52 & -0.84 & 0.62 \\ 
SgrA\textsuperscript{*} & SANE & $+0.94$ & 160 & 50 & -0.07 & 0.54 & -1.53 & 0.57 \\ 
SgrA\textsuperscript{*} & SANE & $+0.94$ & 160 & 70 & -0.89 & 0.56 & -1.22 & 0.50 \\ 
SgrA\textsuperscript{*} & SANE & $+0.94$ & 160 & 90 & 0.01 & 0.64 & 0.01 & 0.46 \\ 
SgrA\textsuperscript{*} & SANE & $+0.94$ & 160 & 110 & 0.91 & 0.53 & 1.19 & 0.44 \\ 
SgrA\textsuperscript{*} & SANE & $+0.94$ & 160 & 130 & 0.06 & 0.51 & 1.56 & 0.52 \\ 
SgrA\textsuperscript{*} & SANE & $+0.94$ & 160 & 150 & -0.44 & 0.44 & 0.92 & 0.56 \\ 
SgrA\textsuperscript{*} & SANE & $+0.94$ & 160 & 170 & 0.36 & 0.84 & -0.23 & 0.97 \\ 
SgrA\textsuperscript{*} & MAD & $-0.94$ & 1 & 10 & 0.37 & 0.40 & -0.15 & 0.25 \\ 
SgrA\textsuperscript{*} & MAD & $-0.94$ & 1 & 30 & 0.33 & 0.35 & -0.18 & 0.21 \\ 
SgrA\textsuperscript{*} & MAD & $-0.94$ & 1 & 50 & 0.28 & 0.26 & -0.20 & 0.19 \\ 
SgrA\textsuperscript{*} & MAD & $-0.94$ & 1 & 70 & 0.25 & 0.19 & -0.09 & 0.23 \\ 
SgrA\textsuperscript{*} & MAD & $-0.94$ & 1 & 90 & 0.01 & 0.15 & -0.05 & 0.26 \\ 
SgrA\textsuperscript{*} & MAD & $-0.94$ & 1 & 110 & -0.21 & 0.17 & 0.06 & 0.23 \\ 
SgrA\textsuperscript{*} & MAD & $-0.94$ & 1 & 130 & -0.32 & 0.19 & 0.23 & 0.19 \\ 
SgrA\textsuperscript{*} & MAD & $-0.94$ & 1 & 150 & -0.46 & 0.26 & 0.29 & 0.21 \\ 
SgrA\textsuperscript{*} & MAD & $-0.94$ & 1 & 170 & -0.55 & 0.32 & 0.32 & 0.24 \\ 
SgrA\textsuperscript{*} & MAD & $-0.94$ & 10 & 10 & 0.73 & 0.57 & -0.08 & 0.37 \\ 
SgrA\textsuperscript{*} & MAD & $-0.94$ & 10 & 30 & 0.68 & 0.50 & -0.09 & 0.36 \\ 
SgrA\textsuperscript{*} & MAD & $-0.94$ & 10 & 50 & 0.63 & 0.39 & -0.03 & 0.35 \\ 
SgrA\textsuperscript{*} & MAD & $-0.94$ & 10 & 70 & 0.62 & 0.34 & 0.16 & 0.39 \\ 
SgrA\textsuperscript{*} & MAD & $-0.94$ & 10 & 90 & 0.03 & 0.31 & -0.05 & 0.40 \\ 
SgrA\textsuperscript{*} & MAD & $-0.94$ & 10 & 110 & -0.60 & 0.28 & -0.18 & 0.42 \\ 
SgrA\textsuperscript{*} & MAD & $-0.94$ & 10 & 130 & -0.74 & 0.30 & 0.07 & 0.34 \\ 
SgrA\textsuperscript{*} & MAD & $-0.94$ & 10 & 150 & -0.88 & 0.41 & 0.21 & 0.34 \\ 
SgrA\textsuperscript{*} & MAD & $-0.94$ & 10 & 170 & -0.98 & 0.48 & 0.25 & 0.35 \\ 
SgrA\textsuperscript{*} & MAD & $-0.94$ & 40 & 10 & 0.92 & 0.63 & -0.04 & 0.50 \\ 
SgrA\textsuperscript{*} & MAD & $-0.94$ & 40 & 30 & 0.91 & 0.58 & -0.00 & 0.50 \\ 
SgrA\textsuperscript{*} & MAD & $-0.94$ & 40 & 50 & 0.88 & 0.46 & 0.09 & 0.48 \\ 
SgrA\textsuperscript{*} & MAD & $-0.94$ & 40 & 70 & 0.72 & 0.38 & 0.19 & 0.44 \\ 
SgrA\textsuperscript{*} & MAD & $-0.94$ & 40 & 90 & 0.01 & 0.38 & 0.04 & 0.36 \\ 
SgrA\textsuperscript{*} & MAD & $-0.94$ & 40 & 110 & -0.71 & 0.35 & -0.09 & 0.41 \\ 
SgrA\textsuperscript{*} & MAD & $-0.94$ & 40 & 130 & -0.99 & 0.41 & 0.03 & 0.43 \\ 
SgrA\textsuperscript{*} & MAD & $-0.94$ & 40 & 150 & -1.13 & 0.53 & 0.17 & 0.43 \\ 
SgrA\textsuperscript{*} & MAD & $-0.94$ & 40 & 170 & -1.19 & 0.58 & 0.26 & 0.42 \\ 
SgrA\textsuperscript{*} & MAD & $-0.94$ & 160 & 10 & 1.01 & 0.60 & -0.16 & 0.57 \\ 
SgrA\textsuperscript{*} & MAD & $-0.94$ & 160 & 30 & 1.00 & 0.57 & -0.10 & 0.60 \\ 
SgrA\textsuperscript{*} & MAD & $-0.94$ & 160 & 50 & 0.92 & 0.48 & -0.02 & 0.58 \\ 
SgrA\textsuperscript{*} & MAD & $-0.94$ & 160 & 70 & 0.67 & 0.44 & 0.06 & 0.45 \\ 
SgrA\textsuperscript{*} & MAD & $-0.94$ & 160 & 90 & 0.01 & 0.43 & 0.08 & 0.38 \\ 
SgrA\textsuperscript{*} & MAD & $-0.94$ & 160 & 110 & -0.67 & 0.40 & 0.16 & 0.39 \\ 
SgrA\textsuperscript{*} & MAD & $-0.94$ & 160 & 130 & -1.00 & 0.44 & 0.28 & 0.48 \\ 
SgrA\textsuperscript{*} & MAD & $-0.94$ & 160 & 150 & -1.16 & 0.54 & 0.37 & 0.53 \\ 
SgrA\textsuperscript{*} & MAD & $-0.94$ & 160 & 170 & -1.22 & 0.57 & 0.43 & 0.51 \\ 
SgrA\textsuperscript{*} & MAD & $-0.5$ & 1 & 10 & 0.42 & 0.38 & -0.23 & 0.25 \\ 
SgrA\textsuperscript{*} & MAD & $-0.5$ & 1 & 30 & 0.35 & 0.29 & -0.23 & 0.17 \\ 
SgrA\textsuperscript{*} & MAD & $-0.5$ & 1 & 50 & 0.23 & 0.18 & -0.21 & 0.15 \\ 
SgrA\textsuperscript{*} & MAD & $-0.5$ & 1 & 70 & 0.17 & 0.17 & -0.09 & 0.21 \\ 
SgrA\textsuperscript{*} & MAD & $-0.5$ & 1 & 90 & -0.01 & 0.17 & 0.04 & 0.24 \\ 
SgrA\textsuperscript{*} & MAD & $-0.5$ & 1 & 110 & -0.19 & 0.16 & 0.12 & 0.18 \\ 
SgrA\textsuperscript{*} & MAD & $-0.5$ & 1 & 130 & -0.23 & 0.18 & 0.21 & 0.13 \\ 
SgrA\textsuperscript{*} & MAD & $-0.5$ & 1 & 150 & -0.29 & 0.29 & 0.19 & 0.17 \\ 
SgrA\textsuperscript{*} & MAD & $-0.5$ & 1 & 170 & -0.33 & 0.36 & 0.17 & 0.25 \\ 
SgrA\textsuperscript{*} & MAD & $-0.5$ & 10 & 10 & 0.90 & 0.55 & -0.11 & 0.35 \\ 
SgrA\textsuperscript{*} & MAD & $-0.5$ & 10 & 30 & 0.80 & 0.44 & -0.09 & 0.27 \\ 
SgrA\textsuperscript{*} & MAD & $-0.5$ & 10 & 50 & 0.69 & 0.32 & -0.01 & 0.28 \\ 
SgrA\textsuperscript{*} & MAD & $-0.5$ & 10 & 70 & 0.64 & 0.29 & 0.20 & 0.35 \\ 
SgrA\textsuperscript{*} & MAD & $-0.5$ & 10 & 90 & -0.06 & 0.28 & 0.01 & 0.40 \\ 
SgrA\textsuperscript{*} & MAD & $-0.5$ & 10 & 110 & -0.67 & 0.28 & -0.18 & 0.33 \\ 
SgrA\textsuperscript{*} & MAD & $-0.5$ & 10 & 130 & -0.67 & 0.34 & 0.02 & 0.25 \\ 
SgrA\textsuperscript{*} & MAD & $-0.5$ & 10 & 150 & -0.72 & 0.48 & 0.07 & 0.26 \\ 
SgrA\textsuperscript{*} & MAD & $-0.5$ & 10 & 170 & -0.78 & 0.58 & 0.06 & 0.32 \\ 
SgrA\textsuperscript{*} & MAD & $-0.5$ & 40 & 10 & 1.23 & 0.56 & 0.03 & 0.45 \\ 
SgrA\textsuperscript{*} & MAD & $-0.5$ & 40 & 30 & 1.20 & 0.47 & 0.09 & 0.38 \\ 
SgrA\textsuperscript{*} & MAD & $-0.5$ & 40 & 50 & 1.11 & 0.38 & 0.18 & 0.37 \\ 
SgrA\textsuperscript{*} & MAD & $-0.5$ & 40 & 70 & 0.85 & 0.36 & 0.22 & 0.39 \\ 
SgrA\textsuperscript{*} & MAD & $-0.5$ & 40 & 90 & -0.09 & 0.34 & -0.03 & 0.41 \\ 
SgrA\textsuperscript{*} & MAD & $-0.5$ & 40 & 110 & -0.91 & 0.35 & -0.24 & 0.41 \\ 
SgrA\textsuperscript{*} & MAD & $-0.5$ & 40 & 130 & -1.10 & 0.45 & -0.17 & 0.35 \\ 
SgrA\textsuperscript{*} & MAD & $-0.5$ & 40 & 150 & -1.11 & 0.58 & -0.08 & 0.37 \\ 
SgrA\textsuperscript{*} & MAD & $-0.5$ & 40 & 170 & -1.12 & 0.65 & -0.04 & 0.39 \\ 
SgrA\textsuperscript{*} & MAD & $-0.5$ & 160 & 10 & 1.23 & 0.52 & -0.11 & 0.53 \\ 
SgrA\textsuperscript{*} & MAD & $-0.5$ & 160 & 30 & 1.21 & 0.49 & -0.03 & 0.52 \\ 
SgrA\textsuperscript{*} & MAD & $-0.5$ & 160 & 50 & 1.08 & 0.40 & 0.01 & 0.46 \\ 
SgrA\textsuperscript{*} & MAD & $-0.5$ & 160 & 70 & 0.75 & 0.36 & 0.01 & 0.41 \\ 
SgrA\textsuperscript{*} & MAD & $-0.5$ & 160 & 90 & -0.07 & 0.38 & -0.02 & 0.38 \\ 
SgrA\textsuperscript{*} & MAD & $-0.5$ & 160 & 110 & -0.81 & 0.37 & -0.07 & 0.37 \\ 
SgrA\textsuperscript{*} & MAD & $-0.5$ & 160 & 130 & -1.07 & 0.43 & -0.07 & 0.46 \\ 
SgrA\textsuperscript{*} & MAD & $-0.5$ & 160 & 150 & -1.15 & 0.52 & -0.05 & 0.52 \\ 
SgrA\textsuperscript{*} & MAD & $-0.5$ & 160 & 170 & -1.17 & 0.55 & 0.01 & 0.49 \\ 
SgrA\textsuperscript{*} & MAD & $0$ & 1 & 10 & 0.42 & 0.34 & -0.20 & 0.26 \\ 
SgrA\textsuperscript{*} & MAD & $0$ & 1 & 30 & 0.32 & 0.27 & -0.18 & 0.19 \\ 
SgrA\textsuperscript{*} & MAD & $0$ & 1 & 50 & 0.13 & 0.17 & -0.18 & 0.12 \\ 
SgrA\textsuperscript{*} & MAD & $0$ & 1 & 70 & -0.01 & 0.15 & -0.12 & 0.20 \\ 
SgrA\textsuperscript{*} & MAD & $0$ & 1 & 90 & -0.06 & 0.21 & 0.09 & 0.28 \\ 
SgrA\textsuperscript{*} & MAD & $0$ & 1 & 110 & -0.07 & 0.16 & 0.24 & 0.21 \\ 
SgrA\textsuperscript{*} & MAD & $0$ & 1 & 130 & -0.06 & 0.14 & 0.24 & 0.13 \\ 
SgrA\textsuperscript{*} & MAD & $0$ & 1 & 150 & -0.11 & 0.22 & 0.11 & 0.16 \\ 
SgrA\textsuperscript{*} & MAD & $0$ & 1 & 170 & -0.13 & 0.30 & 0.04 & 0.24 \\ 
SgrA\textsuperscript{*} & MAD & $0$ & 10 & 10 & 1.00 & 0.53 & -0.01 & 0.38 \\ 
SgrA\textsuperscript{*} & MAD & $0$ & 10 & 30 & 0.82 & 0.47 & -0.03 & 0.30 \\ 
SgrA\textsuperscript{*} & MAD & $0$ & 10 & 50 & 0.51 & 0.35 & -0.09 & 0.28 \\ 
SgrA\textsuperscript{*} & MAD & $0$ & 10 & 70 & 0.34 & 0.29 & 0.04 & 0.42 \\ 
SgrA\textsuperscript{*} & MAD & $0$ & 10 & 90 & -0.06 & 0.37 & 0.09 & 0.45 \\ 
SgrA\textsuperscript{*} & MAD & $0$ & 10 & 110 & -0.39 & 0.28 & 0.08 & 0.40 \\ 
SgrA\textsuperscript{*} & MAD & $0$ & 10 & 130 & -0.34 & 0.30 & 0.20 & 0.27 \\ 
SgrA\textsuperscript{*} & MAD & $0$ & 10 & 150 & -0.50 & 0.40 & 0.01 & 0.26 \\ 
SgrA\textsuperscript{*} & MAD & $0$ & 10 & 170 & -0.62 & 0.50 & -0.14 & 0.35 \\ 
SgrA\textsuperscript{*} & MAD & $0$ & 40 & 10 & 1.47 & 0.63 & 0.10 & 0.56 \\ 
SgrA\textsuperscript{*} & MAD & $0$ & 40 & 30 & 1.34 & 0.60 & 0.13 & 0.45 \\ 
SgrA\textsuperscript{*} & MAD & $0$ & 40 & 50 & 1.09 & 0.44 & 0.09 & 0.38 \\ 
SgrA\textsuperscript{*} & MAD & $0$ & 40 & 70 & 0.78 & 0.39 & 0.12 & 0.44 \\ 
SgrA\textsuperscript{*} & MAD & $0$ & 40 & 90 & -0.06 & 0.42 & 0.03 & 0.40 \\ 
SgrA\textsuperscript{*} & MAD & $0$ & 40 & 110 & -0.86 & 0.38 & -0.08 & 0.43 \\ 
SgrA\textsuperscript{*} & MAD & $0$ & 40 & 130 & -1.01 & 0.44 & -0.08 & 0.36 \\ 
SgrA\textsuperscript{*} & MAD & $0$ & 40 & 150 & -1.09 & 0.59 & -0.25 & 0.40 \\ 
SgrA\textsuperscript{*} & MAD & $0$ & 40 & 170 & -1.12 & 0.66 & -0.37 & 0.54 \\ 
SgrA\textsuperscript{*} & MAD & $0$ & 160 & 10 & 1.42 & 0.38 & -0.17 & 0.54 \\ 
SgrA\textsuperscript{*} & MAD & $0$ & 160 & 30 & 1.34 & 0.40 & -0.13 & 0.54 \\ 
SgrA\textsuperscript{*} & MAD & $0$ & 160 & 50 & 1.10 & 0.35 & -0.15 & 0.44 \\ 
SgrA\textsuperscript{*} & MAD & $0$ & 160 & 70 & 0.74 & 0.31 & -0.11 & 0.36 \\ 
SgrA\textsuperscript{*} & MAD & $0$ & 160 & 90 & -0.10 & 0.37 & -0.08 & 0.36 \\ 
SgrA\textsuperscript{*} & MAD & $0$ & 160 & 110 & -0.90 & 0.33 & -0.14 & 0.36 \\ 
SgrA\textsuperscript{*} & MAD & $0$ & 160 & 130 & -1.15 & 0.35 & -0.13 & 0.44 \\ 
SgrA\textsuperscript{*} & MAD & $0$ & 160 & 150 & -1.28 & 0.43 & -0.29 & 0.61 \\ 
SgrA\textsuperscript{*} & MAD & $0$ & 160 & 170 & -1.29 & 0.44 & -0.27 & 0.61 \\ 
SgrA\textsuperscript{*} & MAD & $+0.5$ & 1 & 10 & 0.30 & 0.34 & -0.22 & 0.22 \\ 
SgrA\textsuperscript{*} & MAD & $+0.5$ & 1 & 30 & 0.22 & 0.26 & -0.21 & 0.15 \\ 
SgrA\textsuperscript{*} & MAD & $+0.5$ & 1 & 50 & 0.07 & 0.15 & -0.20 & 0.12 \\ 
SgrA\textsuperscript{*} & MAD & $+0.5$ & 1 & 70 & -0.05 & 0.18 & -0.14 & 0.22 \\ 
SgrA\textsuperscript{*} & MAD & $+0.5$ & 1 & 90 & -0.08 & 0.23 & 0.10 & 0.33 \\ 
SgrA\textsuperscript{*} & MAD & $+0.5$ & 1 & 110 & -0.05 & 0.18 & 0.29 & 0.24 \\ 
SgrA\textsuperscript{*} & MAD & $+0.5$ & 1 & 130 & -0.02 & 0.15 & 0.27 & 0.16 \\ 
SgrA\textsuperscript{*} & MAD & $+0.5$ & 1 & 150 & -0.07 & 0.24 & 0.16 & 0.14 \\ 
SgrA\textsuperscript{*} & MAD & $+0.5$ & 1 & 170 & -0.08 & 0.32 & 0.09 & 0.21 \\ 
SgrA\textsuperscript{*} & MAD & $+0.5$ & 10 & 10 & 0.55 & 0.64 & -0.25 & 0.26 \\ 
SgrA\textsuperscript{*} & MAD & $+0.5$ & 10 & 30 & 0.37 & 0.53 & -0.30 & 0.22 \\ 
SgrA\textsuperscript{*} & MAD & $+0.5$ & 10 & 50 & 0.07 & 0.38 & -0.38 & 0.31 \\ 
SgrA\textsuperscript{*} & MAD & $+0.5$ & 10 & 70 & -0.07 & 0.32 & -0.26 & 0.46 \\ 
SgrA\textsuperscript{*} & MAD & $+0.5$ & 10 & 90 & -0.11 & 0.37 & 0.13 & 0.56 \\ 
SgrA\textsuperscript{*} & MAD & $+0.5$ & 10 & 110 & -0.01 & 0.28 & 0.48 & 0.49 \\ 
SgrA\textsuperscript{*} & MAD & $+0.5$ & 10 & 130 & 0.07 & 0.38 & 0.53 & 0.39 \\ 
SgrA\textsuperscript{*} & MAD & $+0.5$ & 10 & 150 & -0.08 & 0.54 & 0.29 & 0.28 \\ 
SgrA\textsuperscript{*} & MAD & $+0.5$ & 10 & 170 & -0.18 & 0.64 & 0.12 & 0.29 \\ 
SgrA\textsuperscript{*} & MAD & $+0.5$ & 40 & 10 & 0.88 & 1.06 & -0.37 & 0.48 \\ 
SgrA\textsuperscript{*} & MAD & $+0.5$ & 40 & 30 & 0.73 & 0.89 & -0.45 & 0.40 \\ 
SgrA\textsuperscript{*} & MAD & $+0.5$ & 40 & 50 & 0.43 & 0.59 & -0.52 & 0.63 \\ 
SgrA\textsuperscript{*} & MAD & $+0.5$ & 40 & 70 & 0.20 & 0.50 & -0.32 & 0.79 \\ 
SgrA\textsuperscript{*} & MAD & $+0.5$ & 40 & 90 & -0.06 & 0.43 & 0.14 & 0.70 \\ 
SgrA\textsuperscript{*} & MAD & $+0.5$ & 40 & 110 & -0.28 & 0.52 & 0.70 & 0.78 \\ 
SgrA\textsuperscript{*} & MAD & $+0.5$ & 40 & 130 & -0.23 & 0.60 & 0.78 & 0.68 \\ 
SgrA\textsuperscript{*} & MAD & $+0.5$ & 40 & 150 & -0.22 & 0.87 & 0.44 & 0.44 \\ 
SgrA\textsuperscript{*} & MAD & $+0.5$ & 40 & 170 & -0.25 & 1.06 & 0.11 & 0.54 \\ 
SgrA\textsuperscript{*} & MAD & $+0.5$ & 160 & 10 & 0.77 & 0.94 & -0.41 & 0.82 \\ 
SgrA\textsuperscript{*} & MAD & $+0.5$ & 160 & 30 & 0.83 & 0.83 & -0.47 & 0.67 \\ 
SgrA\textsuperscript{*} & MAD & $+0.5$ & 160 & 50 & 0.65 & 0.50 & -0.58 & 0.45 \\ 
SgrA\textsuperscript{*} & MAD & $+0.5$ & 160 & 70 & 0.26 & 0.48 & -0.49 & 0.55 \\ 
SgrA\textsuperscript{*} & MAD & $+0.5$ & 160 & 90 & -0.07 & 0.42 & 0.06 & 0.49 \\ 
SgrA\textsuperscript{*} & MAD & $+0.5$ & 160 & 110 & -0.31 & 0.46 & 0.62 & 0.50 \\ 
SgrA\textsuperscript{*} & MAD & $+0.5$ & 160 & 130 & -0.52 & 0.50 & 0.65 & 0.49 \\ 
SgrA\textsuperscript{*} & MAD & $+0.5$ & 160 & 150 & -0.42 & 0.75 & 0.25 & 0.67 \\ 
SgrA\textsuperscript{*} & MAD & $+0.5$ & 160 & 170 & -0.27 & 0.95 & -0.01 & 0.91 \\ 
SgrA\textsuperscript{*} & MAD & $+0.94$ & 1 & 10 & 0.31 & 0.18 & -0.29 & 0.14 \\ 
SgrA\textsuperscript{*} & MAD & $+0.94$ & 1 & 30 & 0.26 & 0.15 & -0.28 & 0.12 \\ 
SgrA\textsuperscript{*} & MAD & $+0.94$ & 1 & 50 & 0.19 & 0.12 & -0.27 & 0.10 \\ 
SgrA\textsuperscript{*} & MAD & $+0.94$ & 1 & 70 & 0.14 & 0.12 & -0.24 & 0.13 \\ 
SgrA\textsuperscript{*} & MAD & $+0.94$ & 1 & 90 & 0.03 & 0.14 & -0.03 & 0.17 \\ 
SgrA\textsuperscript{*} & MAD & $+0.94$ & 1 & 110 & -0.09 & 0.12 & 0.19 & 0.13 \\ 
SgrA\textsuperscript{*} & MAD & $+0.94$ & 1 & 130 & -0.16 & 0.11 & 0.25 & 0.09 \\ 
SgrA\textsuperscript{*} & MAD & $+0.94$ & 1 & 150 & -0.27 & 0.13 & 0.28 & 0.10 \\ 
SgrA\textsuperscript{*} & MAD & $+0.94$ & 1 & 170 & -0.34 & 0.15 & 0.31 & 0.12 \\ 
SgrA\textsuperscript{*} & MAD & $+0.94$ & 10 & 10 & 0.37 & 0.27 & -0.39 & 0.17 \\ 
SgrA\textsuperscript{*} & MAD & $+0.94$ & 10 & 30 & 0.27 & 0.25 & -0.40 & 0.15 \\ 
SgrA\textsuperscript{*} & MAD & $+0.94$ & 10 & 50 & 0.13 & 0.21 & -0.43 & 0.16 \\ 
SgrA\textsuperscript{*} & MAD & $+0.94$ & 10 & 70 & 0.10 & 0.20 & -0.36 & 0.23 \\ 
SgrA\textsuperscript{*} & MAD & $+0.94$ & 10 & 90 & 0.01 & 0.21 & -0.04 & 0.26 \\ 
SgrA\textsuperscript{*} & MAD & $+0.94$ & 10 & 110 & -0.06 & 0.19 & 0.31 & 0.22 \\ 
SgrA\textsuperscript{*} & MAD & $+0.94$ & 10 & 130 & -0.12 & 0.20 & 0.39 & 0.17 \\ 
SgrA\textsuperscript{*} & MAD & $+0.94$ & 10 & 150 & -0.29 & 0.22 & 0.38 & 0.14 \\ 
SgrA\textsuperscript{*} & MAD & $+0.94$ & 10 & 170 & -0.40 & 0.24 & 0.39 & 0.16 \\ 
SgrA\textsuperscript{*} & MAD & $+0.94$ & 40 & 10 & 0.37 & 0.42 & -0.62 & 0.23 \\ 
SgrA\textsuperscript{*} & MAD & $+0.94$ & 40 & 30 & 0.24 & 0.40 & -0.67 & 0.24 \\ 
SgrA\textsuperscript{*} & MAD & $+0.94$ & 40 & 50 & 0.07 & 0.38 & -0.76 & 0.32 \\ 
SgrA\textsuperscript{*} & MAD & $+0.94$ & 40 & 70 & 0.10 & 0.35 & -0.58 & 0.40 \\ 
SgrA\textsuperscript{*} & MAD & $+0.94$ & 40 & 90 & 0.01 & 0.29 & -0.02 & 0.37 \\ 
SgrA\textsuperscript{*} & MAD & $+0.94$ & 40 & 110 & -0.03 & 0.32 & 0.53 & 0.39 \\ 
SgrA\textsuperscript{*} & MAD & $+0.94$ & 40 & 130 & -0.05 & 0.34 & 0.69 & 0.33 \\ 
SgrA\textsuperscript{*} & MAD & $+0.94$ & 40 & 150 & -0.25 & 0.36 & 0.62 & 0.23 \\ 
SgrA\textsuperscript{*} & MAD & $+0.94$ & 40 & 170 & -0.40 & 0.38 & 0.59 & 0.20 \\ 
SgrA\textsuperscript{*} & MAD & $+0.94$ & 160 & 10 & 0.39 & 0.51 & -0.72 & 0.33 \\ 
SgrA\textsuperscript{*} & MAD & $+0.94$ & 160 & 30 & 0.30 & 0.49 & -0.82 & 0.31 \\ 
SgrA\textsuperscript{*} & MAD & $+0.94$ & 160 & 50 & 0.19 & 0.41 & -0.91 & 0.38 \\ 
SgrA\textsuperscript{*} & MAD & $+0.94$ & 160 & 70 & 0.08 & 0.38 & -0.70 & 0.45 \\ 
SgrA\textsuperscript{*} & MAD & $+0.94$ & 160 & 90 & -0.01 & 0.31 & -0.00 & 0.43 \\ 
SgrA\textsuperscript{*} & MAD & $+0.94$ & 160 & 110 & -0.02 & 0.36 & 0.70 & 0.39 \\ 
SgrA\textsuperscript{*} & MAD & $+0.94$ & 160 & 130 & -0.14 & 0.41 & 0.89 & 0.33 \\ 
SgrA\textsuperscript{*} & MAD & $+0.94$ & 160 & 150 & -0.28 & 0.48 & 0.82 & 0.29 \\ 
SgrA\textsuperscript{*} & MAD & $+0.94$ & 160 & 170 & -0.38 & 0.51 & 0.72 & 0.35 \\ 
M87\textsuperscript{*} & SANE & $-0.94$ & 1 & 17 & -0.22 & 3.58 & 0.34 & 3.53 \\ 
M87\textsuperscript{*} & SANE & $-0.94$ & 10 & 17 & 0.02 & 1.10 & 0.17 & 0.97 \\ 
M87\textsuperscript{*} & SANE & $-0.94$ & 20 & 17 & 0.21 & 0.73 & 0.13 & 0.53 \\ 
M87\textsuperscript{*} & SANE & $-0.94$ & 40 & 17 & 0.32 & 0.48 & 0.15 & 0.34 \\ 
M87\textsuperscript{*} & SANE & $-0.94$ & 80 & 17 & 0.43 & 0.38 & 0.19 & 0.31 \\ 
M87\textsuperscript{*} & SANE & $-0.94$ & 160 & 17 & 0.62 & 0.44 & 0.27 & 0.52 \\ 
M87\textsuperscript{*} & SANE & $-0.5$ & 1 & 17 & -0.71 & 4.54 & 0.89 & 4.48 \\ 
M87\textsuperscript{*} & SANE & $-0.5$ & 10 & 17 & -0.04 & 1.20 & 0.39 & 1.21 \\ 
M87\textsuperscript{*} & SANE & $-0.5$ & 20 & 17 & 0.36 & 0.77 & 0.59 & 0.79 \\ 
M87\textsuperscript{*} & SANE & $-0.5$ & 40 & 17 & 0.67 & 0.55 & 0.65 & 0.66 \\ 
M87\textsuperscript{*} & SANE & $-0.5$ & 80 & 17 & 0.82 & 0.44 & 0.69 & 0.61 \\ 
M87\textsuperscript{*} & SANE & $-0.5$ & 160 & 17 & 0.89 & 0.46 & 0.67 & 0.66 \\ 
M87\textsuperscript{*} & SANE & $0$ & 1 & 17 & -1.43 & 3.32 & 0.65 & 3.23 \\ 
M87\textsuperscript{*} & SANE & $0$ & 10 & 17 & -2.93 & 3.71 & 3.66 & 3.22 \\ 
M87\textsuperscript{*} & SANE & $0$ & 20 & 17 & -0.89 & 3.43 & 3.34 & 2.27 \\ 
M87\textsuperscript{*} & SANE & $0$ & 40 & 17 & 1.11 & 2.70 & 2.65 & 1.51 \\ 
M87\textsuperscript{*} & SANE & $0$ & 80 & 17 & 2.45 & 1.97 & 2.30 & 1.13 \\ 
M87\textsuperscript{*} & SANE & $0$ & 160 & 17 & 3.20 & 1.63 & 2.37 & 1.24 \\ 
M87\textsuperscript{*} & SANE & $+0.5$ & 1 & 163 & 0.58 & 0.91 & 0.35 & 0.99 \\ 
M87\textsuperscript{*} & SANE & $+0.5$ & 10 & 163 & -1.98 & 3.11 & 2.00 & 3.08 \\ 
M87\textsuperscript{*} & SANE & $+0.5$ & 20 & 163 & -1.96 & 2.21 & 1.74 & 2.16 \\ 
M87\textsuperscript{*} & SANE & $+0.5$ & 40 & 163 & -0.93 & 1.48 & -0.42 & 1.86 \\ 
M87\textsuperscript{*} & SANE & $+0.5$ & 80 & 163 & -0.57 & 1.41 & -1.95 & 1.58 \\ 
M87\textsuperscript{*} & SANE & $+0.5$ & 160 & 163 & -0.76 & 1.41 & -2.76 & 1.38 \\ 
M87\textsuperscript{*} & SANE & $+0.94$ & 1 & 163 & 0.47 & 0.22 & 0.44 & 0.17 \\ 
M87\textsuperscript{*} & SANE & $+0.94$ & 10 & 163 & 1.81 & 4.79 & -0.07 & 4.70 \\ 
M87\textsuperscript{*} & SANE & $+0.94$ & 20 & 163 & 0.94 & 2.13 & -0.87 & 2.24 \\ 
M87\textsuperscript{*} & SANE & $+0.94$ & 40 & 163 & 1.15 & 1.61 & -1.08 & 1.90 \\ 
M87\textsuperscript{*} & SANE & $+0.94$ & 80 & 163 & 1.08 & 1.23 & -1.00 & 1.58 \\ 
M87\textsuperscript{*} & SANE & $+0.94$ & 160 & 163 & 0.99 & 0.96 & -0.78 & 1.36 \\ 
M87\textsuperscript{*} & MAD & $-0.94$ & 1 & 17 & 0.25 & 0.15 & 0.00 & 0.15 \\ 
M87\textsuperscript{*} & MAD & $-0.94$ & 10 & 17 & 0.33 & 0.22 & 0.05 & 0.23 \\ 
M87\textsuperscript{*} & MAD & $-0.94$ & 20 & 17 & 0.33 & 0.25 & 0.06 & 0.26 \\ 
M87\textsuperscript{*} & MAD & $-0.94$ & 40 & 17 & 0.31 & 0.28 & 0.07 & 0.29 \\ 
M87\textsuperscript{*} & MAD & $-0.94$ & 80 & 17 & 0.33 & 0.33 & 0.10 & 0.34 \\ 
M87\textsuperscript{*} & MAD & $-0.94$ & 160 & 17 & 0.42 & 0.44 & 0.24 & 0.51 \\ 
M87\textsuperscript{*} & MAD & $-0.5$ & 1 & 17 & 0.35 & 0.19 & 0.13 & 0.14 \\ 
M87\textsuperscript{*} & MAD & $-0.5$ & 10 & 17 & 0.68 & 0.24 & 0.42 & 0.27 \\ 
M87\textsuperscript{*} & MAD & $-0.5$ & 20 & 17 & 0.73 & 0.26 & 0.48 & 0.31 \\ 
M87\textsuperscript{*} & MAD & $-0.5$ & 40 & 17 & 0.67 & 0.29 & 0.42 & 0.36 \\ 
M87\textsuperscript{*} & MAD & $-0.5$ & 80 & 17 & 0.58 & 0.29 & 0.37 & 0.42 \\ 
M87\textsuperscript{*} & MAD & $-0.5$ & 160 & 17 & 0.60 & 0.31 & 0.48 & 0.56 \\ 
M87\textsuperscript{*} & MAD & $0$ & 1 & 17 & 0.28 & 0.24 & 0.12 & 0.13 \\ 
M87\textsuperscript{*} & MAD & $0$ & 10 & 17 & 0.83 & 0.39 & 0.60 & 0.32 \\ 
M87\textsuperscript{*} & MAD & $0$ & 20 & 17 & 1.01 & 0.45 & 0.73 & 0.39 \\ 
M87\textsuperscript{*} & MAD & $0$ & 40 & 17 & 0.93 & 0.42 & 0.58 & 0.42 \\ 
M87\textsuperscript{*} & MAD & $0$ & 80 & 17 & 0.76 & 0.29 & 0.43 & 0.46 \\ 
M87\textsuperscript{*} & MAD & $0$ & 160 & 17 & 0.74 & 0.29 & 0.51 & 0.54 \\ 
M87\textsuperscript{*} & MAD & $+0.5$ & 1 & 163 & 0.04 & 0.26 & 0.06 & 0.11 \\ 
M87\textsuperscript{*} & MAD & $+0.5$ & 10 & 163 & 0.07 & 0.60 & 0.09 & 0.28 \\ 
M87\textsuperscript{*} & MAD & $+0.5$ & 20 & 163 & 0.05 & 0.78 & 0.04 & 0.33 \\ 
M87\textsuperscript{*} & MAD & $+0.5$ & 40 & 163 & 0.02 & 0.84 & -0.14 & 0.46 \\ 
M87\textsuperscript{*} & MAD & $+0.5$ & 80 & 163 & 0.04 & 0.75 & -0.31 & 0.60 \\ 
M87\textsuperscript{*} & MAD & $+0.5$ & 160 & 163 & 0.12 & 0.72 & -0.48 & 0.77 \\ 
M87\textsuperscript{*} & MAD & $+0.94$ & 1 & 163 & -0.07 & 0.11 & 0.13 & 0.06 \\ 
M87\textsuperscript{*} & MAD & $+0.94$ & 10 & 163 & 0.02 & 0.23 & 0.26 & 0.14 \\ 
M87\textsuperscript{*} & MAD & $+0.94$ & 20 & 163 & 0.07 & 0.29 & 0.35 & 0.19 \\ 
M87\textsuperscript{*} & MAD & $+0.94$ & 40 & 163 & 0.09 & 0.36 & 0.45 & 0.23 \\ 
M87\textsuperscript{*} & MAD & $+0.94$ & 80 & 163 & 0.08 & 0.44 & 0.45 & 0.26 \\ 
M87\textsuperscript{*} & MAD & $+0.94$ & 160 & 163 & 0.13 & 0.48 & 0.37 & 0.33 \\ 

\end{longtable*}

\newpage
\bibliographystyle{aasjournal}
\bibliography{references}

\end{document}